\documentclass[12pt,letterpaper,oneside]{article}
\usepackage{amssymb,tabularx,multirow,amsmath,natbib, epsfig, threeparttable, amstext, subfigure,xcolor}
\usepackage{pdflscape}
\usepackage{afterpage}
\usepackage{capt-of}
\usepackage{rotating}

\newcommand{\rf}{\vskip .1in\par\sloppy\hangindent=1pc\hangafter=1
                 \noindent}

\newcommand{\bone}[1]{\mathbf{1}_{\lbr #1 \rbr}}
\newcommand{\lb}{\left[}
\newcommand{\rb}{\right]}
\newcommand{\lbr}{\left\{}
\newcommand{\rbr}{\right\}}
\newcommand{\lp}{\left(}
\newcommand{\rp}{\right)}

\usepackage[dvips,top=1.2in,bottom=0.65in,left=1.15in,right=1.15in,includefoot]{geometry}

\usepackage[margin=20pt,labelfont={bf,small},textfont={it,small},indention=0.5cm]{caption}

\begin{document}

\def\spacingset#1{\renewcommand{\baselinestretch}%
{#1}\small\normalsize} \spacingset{1}


\title{\bf Running on empty: Recharge dynamics from animal movement data}
\author{Mevin B. Hooten \\
  \small U.S. Geological Survey, Colorado Cooperative Fish and Wildlife Research Unit \\ \small Department of Fish, Wildlife, and Conservation and Department of Statistics \\ \small Colorado State University\\ \small mevin.hooten@colostate.edu
  \\
   Henry R. Scharf \\ \small Department of Statistics, Colorado State University \\ \small henry.scharf@colostate.edu
  \\
   Juan M. Morales \\ \small Grupo de Ecolog\'ia Cuantitativa, INIBIOMA, Universidad Nacional del Comahue, CONICET \\ \small jm.morales@conicet.gov.ar}
\maketitle

\noindent \textbf{Corresponding Author}:  Mevin B. Hooten, 1484 Campus Delivery, Colorado State University, Fort Collins, CO  80523.  Phone:  970-491-1415, Email: mevin.hooten@colostate.edu

\bigskip
\pagebreak
\begin{abstract}
Vital rates such as survival and recruitment have always been important in the study of population and community ecology.  At the individual level, physiological processes such as energetics are critical in understanding biomechanics and movement ecology and also scale up to influence food webs and trophic cascades.  Although vital rates and population-level characteristics are tied with individual-level animal movement, most statistical models for telemetry data are not equipped to provide inference about these relationships because they lack the explicit, mechanistic connection to physiological dynamics.  We present a framework for modeling telemetry data that explicitly includes an aggregated physiological process associated with decision making and movement in heterogeneous environments.  Our framework accommodates a wide range of movement and physiological process specifications.  We illustrate a specific model formulation in continuous-time to provide direct inference about gains and losses associated with physiological processes based on movement.  Our approach can also be extended to accommodate auxiliary data when available.  We demonstrate our model to infer mountain lion (\emph{Puma concolor}; in Colorado, USA) and African buffalo (\emph{Syncerus caffer}; in Kruger National Park, South Africa) recharge dynamics.
\end{abstract}

\vfill

\newpage

\spacingset{1.45} 

\section{Introduction}
Energetics has been a dominant theme in ecological and biological science for centuries (Zuntz, 1897; Nussbaum, 1978) because an improved understanding of metabolics and energy acquisition provides insights about fundamental similarities and differences among species (Taylor et al., 1982).  An understanding of the connection between energetics and movement is critical for all aspects of biology and leads to improved management and conservation of wildlife because physiological processes and vital rates are indicative of animal health (Nathan et al., 2008; Wilmers et al., 2017).  Healthy wildlife individuals and populations are an essential ecosystem service and have intrinsic anthropogenic and ecosystem value (Ingraham and Foster, 2008).

While much research has focused primarily on the ties between energy and locomotion, myriad other factors influence animal decision making processes (Alcock, 2009).  Decisions made by animals directly affect their movement rates and hence indirectly affect their energy as well as other physiological processes (Houston and McNamara, 1999; Morales et al., 2005, 2010).  In what follows, we use the term ``recharge'' as a general reference to physiological processes that require replenishment for an organism to maintain its physical health and normal activities.  The recharge concept is a simplification of complex physiological changes over time; it reduces the complexity enough that we can account for aggregate physiological signals while inferring environmental influences on animal movement given telemetry data.  We describe examples of physiological processes that may be connected with animal movement decisions and show how they accumulate in a recharge function that can be statistically inferred using tracking data.  Our approach to account for recharge dynamics relies on a long-memory statistical model specified to mimic physiological processes and can be applied to animal tracking data to test hypotheses about animal behavior as well as estimate parameters associated with changes in physiological processes over time.    

Many former studies of animal movement have used experimental laboratory approaches to measure oxygen intake and energy expenditure directly (Alexander, 2003; Halsey, 2016).  These studies provided a foundational kinematic understanding of animal movement in controlled environments (Full et al., 1990).  More recent research has examined connections between movement and energetics in natural settings (Karasov, 1992) and how terrain and environmental factors influence movement (e.g., Humphries and Careau, 2011; Shepard et al., 2013; Williams et al., 2014).  Biotelemetry technology has facilitated regular measurement of movement and led to improved understanding of individual-based physiological processes (e.g., Cooke et al., 2004; Green, 2011).   

Improvements in high-quality animal tracking data are occurring at an increasing rate (Cagnacci et al., 2010).  Wildlife tracking devices have allowed researchers to collect unprecedented data sets that contain valuable information about animal movement, and hence energetics and other physiological processes that require recharge (Kays et al., 2015; Wilmers et al., 2015).   Statistical approaches have been developed to characterize the variation within and among individual animal trajectories (Scharf et al., 2016; Hooten and Johnson, 2019; Hooten et al., 2017; Scharf et al., 2018).  These approaches include the use of environmental information and methods to identify the portions of animal trajectories that indicate distinctly different patterns (e.g., Whoriskey et al., 2017).  For example, stochastic differential equations (SDEs; Brillinger, 2010) allow researchers to make inference on the importance of environmental covariates on movement in continuous time.  Some discrete-time models also incorporate covariates and focus on phenomenological clustering of movement processes that are linked to possible behavioral changes over time (e.g., Morales et al., 2004; Langrock et al., 2012; McClintock et al., 2012; McKellar et al., 2015).  

Despite the proliferation of statistical animal movement models, few are based on specific mechanisms related to physiology (e.g., Schick et al., 2013).  By contrast, purely mathematical animal movement models are almost always mechanistically motivated (Turchin, 1998), but are often too complex to allow for statistical learning using location-based telemetry data alone.  Some statistical models have been used to make \emph{post hoc} inferences concerning physiological processes such as memory (e.g., Avgar et al., 2013; Oliveira-Santos et al., 2016) and energetics (e.g., Merkle et al., 2017; Hooten et al., 2018), including some that rely on auxiliary data from accelerometers (e.g., Wilson et al., 2012).  However, they often lack the mechanistic mathematical specifications to account for recharge dynamics directly when inferring movement dynamics.  Demographic models based on capture-recapture data, such as Cormack-Jolly-Seber (CJS) models, explicitly consider individual health and body condition when inferring vital rates (Lebreton et al., 1992; Pollock, 1991), but are often focused on large spatial and temporal scales (Schick et al., 2013).   

In what follows, we broaden the current scope of ``energy landscapes'' (Wilson et al., 2012; Shepard et al., 2013) and ``landscapes of fear'' (Laundr{\'e} et al., 2001; Bleicher, 2017) to include all physiological processes that require recharge.  We consider accumulations of these physiological landscapes that result in individual-based recharge functions and link them to decision making processes of individual organisms as they move.  We show how to use telemetry data to make inference about both the decision and recharge processes in heterogeneous environments and account for their effect on movement.  We demonstrate our recharge movement model with case studies involving telemetry data for a mountain lion (\emph{Puma concolor}) in Colorado, USA and African buffalo (\emph{Syncerus caffer}) in South Africa.  We also discuss possible ways to extend the model to account for conspecific and allospecific interactions among individuals as well as accommodate auxiliary data sources such as individual-level health and accelerometer data.    

\section{Material and Methods}
\subsection{Physiological Landscapes}
Critical to our approach is the concept of recharge, a time-varying process involving an individual physiological characteristic $v$.  Physiological recharge can be expressed as a function $g(v,t)$ that increases (i.e., charges) and decreases (i.e., discharges) over time depending on the decision making process of the individual, the resulting behavior, and the environmental conditions it encounters.  We refer to a combination of spatially explicit covariates that affect the recharge function $g(v,t)$ over time as the ``physiological landscape.''  For a physiological characteristic $v$, we define the physiological landscape as $\mathbf{w}'(\boldsymbol\mu)\boldsymbol\theta(v)$ for any location $\boldsymbol\mu\equiv(\mu_1,\mu_2)'$ in region ${\cal D}$ (e.g., the study area).  

\begin{figure}[htp]
\centering
\includegraphics[width=4in]{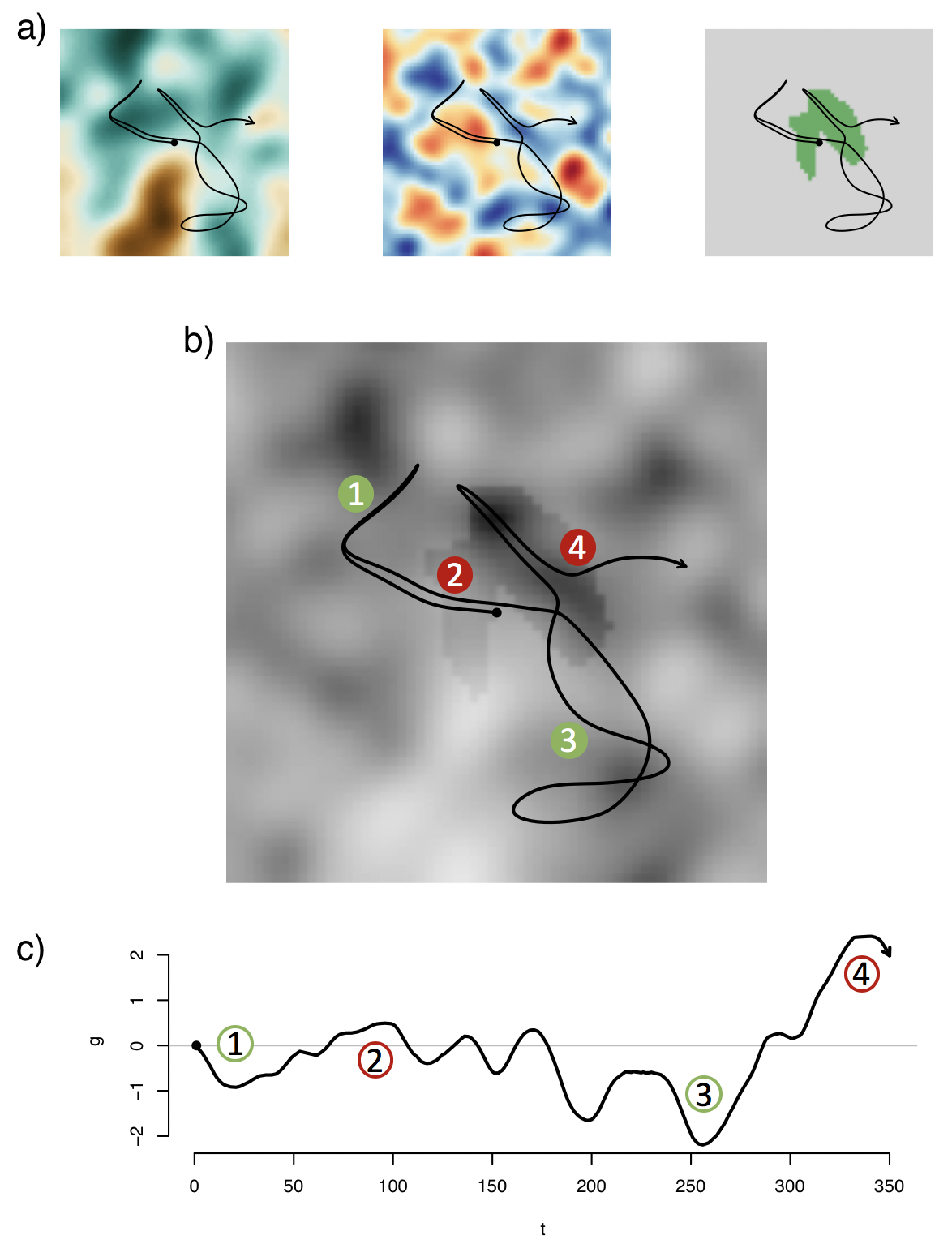}
\caption{a) The simulated environmental covariates $\mathbf{w}(\boldsymbol\mu)$ that may influence the recharge function (left: large scale spatial process, middle:  small scale spatial process, right: patch; b) an example physiological landscape based on the environmental covariates with example individual trajectory ($\boldsymbol\mu(t)$, for all $t\in {\cal T}$) shown as solid line beginning at solid point and ending at the arrow; c) The physiological recharge function arising from the path integral of the physiological landscape associated with trajectory.  Numbered circles represent time points at which the simulated individual is charged (red) and discharged (green).}
\label{fig:gvt}
\end{figure}

The coefficients $\boldsymbol\theta(v)\equiv(\theta_1(v),\ldots,\theta_p(v))'$ appropriately weight each of the landscape variables (e.g., elevation, land type, etc.) in $\mathbf{w}(\boldsymbol\mu)\equiv(w_1(\boldsymbol\mu),\ldots,w_p(\boldsymbol\mu))'$ so that they combine to result in a surface that modulates the state of the physiological process $v$ as an individual moves throughout the space (Figure~\ref{fig:gvt}).  For example, if $v$ refers to the energetic component of a larger suite of physiological processes, then $\mathbf{w}'(\boldsymbol\mu(t))\boldsymbol\theta(v)$ represents the physiological landscape value that influences the energy recharge dynamics as the individual under study moves to position $\boldsymbol\mu(t)$ at time $t$.    

In fact, for a given individual trajectory $\boldsymbol\mu(t)$ (for all $t\in {\cal T}$ in the study period), the physiological landscape $\mathbf{w}'(\boldsymbol\mu(t))\boldsymbol\theta(v)$ is accumulated as the individual moves.  This accumulation over time results in what we refer to as a physiological recharge function that can be expressed as the line integral of the physiological landscape 
\begin{equation}    
  g(v,t)=g_0(v)+\int_0^t \mathbf{w}'(\boldsymbol\mu(\tau))\boldsymbol\theta(v) d\tau \,,
  \label{eq:gvt}
\end{equation}    
\noindent where the lower limit (i.e., zero) on the integral in (\ref{eq:gvt}) corresponds to the beginning of the study period.  Figure~\ref{fig:gvt}c depicts the physiological recharge function as the line integral associated with the trajectory.  At times when $g(v,t)$ is large, the individual is in a charged state with respect to physiological process $v$.  Conversely, when the physiological recharge function $g(v,t)$ is small, it indicates that the physiological process $v$ is discharged and the individual may alter its behavior in an attempt to recharge.   

While energy is among the most commonly studied physiological characteristic (Wilson et al., 2012), there exists a large set of other individual-based physiological characteristics (i.e., $v\in {\cal V}$) that contribute to individual, population, community, and ecosystem health and larger scale vital rates (Matthiopoulos et al., 2015).  For example, in addition to energy intake and expenditure (Spalinger and Hobbs, 1992; Stephens et al., 2007), most animals require periodic hydration (e.g., Tshipa et al., 2017), sleep (Savage and West, 2007), heat (Humphries and Careau, 2011), and shelter (Eggleston and Lipcius, 1992).  Less obvious physiological processes requiring recharge that transcend the individual level may include activities such as reproduction (Proaktor et al., 2008), care for young (Dudek et al., 2018), and ``security'' in the context of landscapes of fear (Laundr{\'e} et al., 2001; Bleicher, 2017).  Thus, we can express an aggregated physiological recharge process as an integral over the set of all physiological processes ${\cal V}$:

\begin{align}
  g(t) &= \int_{{\cal V}} g(v,t) dv \,, \\ 
  &= g_0 + \int_0^t \mathbf{w}'(\boldsymbol\mu(\tau))\boldsymbol\theta d\tau \,, \label{eq:gt} 
\end{align}
\noindent where we show that the initial aggregated charge is 
\begin{equation} 
  g_0\equiv\int_{{\cal V}} g_0(v) dv \,, 
\end{equation} 
and the aggregated recharge coefficients are  
\begin{equation} 
  \boldsymbol\theta\equiv\int_{{\cal V}} \boldsymbol\theta(v) dv 
\end{equation} 
\noindent in Online Appendix A.  As we describe in what follows, the aggregated recharge process in (\ref{eq:gt}) provides a fundamental mechanistic link between environmental characteristics and the physiology and sociality of moving individuals as they seek to recharge --- a link that is missing in most other contemporary models for animal movement and one that allows us to examine the evidence for physiological signals in animal movement trajectories.  Furthermore, in the absence of a strict connection to specific physiological processes, the recharge function in (\ref{eq:gt}) can be used to generalize movement models to accommodate long-range temporal dependence that may go unaccounted for otherwise.  Finally, the recharge function we specified in (\ref{eq:gt}) can be generalized easily to accommodate time varying coefficients (i.e., $\boldsymbol\theta(t)$), nonlinearity in the physiological landscape, and alternatives to the convolution form of aggregation (e.g., based on the principle of limiting factors).  For example, to account for optima in the environmental gradients that comprise the physiological landscape, we can include polynomial transformations of environmental variables $\mathbf{w}$ as we would in a conventional regression model.       

\subsection{Movement Decisions Based on Physiological Processes}
\subsubsection{General Framework}
Most modern statistical models for animal trajectories account for both measurement error and movement dynamics using a hierarchical framework (Schick et al., 2008; Hooten et al., 2017).  Thus, we employ a hierarchical structure to build a general modeling framework that reconciles animal trajectories and physiological processes while accounting for measurement error and uncertainty in movement dynamics (Figure~\ref{fig:model}).  To develop a general recharge-based movement modeling framework, we consider a model for the telemetry data that depends on the true, underlying animal trajectory.  Our movement model characterizes the structure of the trajectory, and hence the perception of the landscape by the animal, depending on a binary decision process $\mathbf{z}(t_i)$ of the animal over time.  This decision process arises stochastically according to a probability function that depends on the underlying aggregated physiological process.  For telemetry observations $\mathbf{s}(t_i)$ (for $i=1,\ldots,n$) and associated trajectory $\boldsymbol\mu(t_i)$ we formulate the hierarchical model    

\begin{align}
  \mathbf{s}(t_i) &\sim [\mathbf{s}(t_i) | \boldsymbol\mu(t_i)] \,, \label{eq:data_model} \\
  \boldsymbol\mu(t_i) &\sim  
  \begin{cases}
    {\cal M}_0 &\mbox{, } z(t_i)=0 \,, \\
    {\cal M}_1 &\mbox{, } z(t_i)=1 \,, 
  \end{cases} \label{eq:process_model}
\end{align}
\noindent for $i=1,\ldots,n$, where the bracket notation `$[\cdot]$' denotes a generic probability distribution (Gelfand and Smith, 1990) that may include additional parameters.  We introduce continuous-time models for ${\cal M}_0$ and ${\cal M}_1$ in the example specification that follows.  

The mixture movement model in (\ref{eq:process_model}) depends on a latent binary decision $z(t)$ that represents the individual's choice to recharge when $z(t)=1$ (where $z(t)=1$ corresponds to a discharged state and $z(t)=0$ corresponds to a charged state).  The instantaneous probability of the decision to recharge ($\rho(t)$) can be related to the latent physiological recharge process $g(t)$ through an appropriate link function.  Thus, in the case studies that follow, we express $z(t)\sim\text{Bern}(\rho(t))$ with $\rho(t)=1-\Phi(g(t))$, where $\Phi(\cdot)$ represents the standard normal cumulative distribution function (i.e., the inverse probit function; another option is the logit).  This relationship between $\rho(t)$ and $g(t)$ implies that the decision to recharge will increase in probability when the aggregated physiological process $g(t)$ decreases.  For example, as an individual ventures far from resources that allow it to recharge, $g(t)$ will decline and the individual will eventually need to make an effort to replenish its physiological processes, hence increasing $\rho(t)$ and changing its movement behavior (Figure~\ref{fig:model}). By connecting an animal decision process $z(t)$ with movement and resources, our model formulation explicitly accounts for the relationship between stimuli and motivation, which is a primary focus of ethology (Colgan, 1989).  

\subsubsection{A Continuous-Time Recharge-Based Movement Model}
In the continuous-time setting, stochastic differential equation (SDE) models are a popular option to account for diffusion and drift across heterogeneous landscapes (Brillinger, 2010).  Thus, we provide an example specification using the hierarchical framework by formulating the specific components of our recharge-based movement model in (\ref{eq:data_model})--(\ref{eq:process_model}).  We consider Gaussian error for telemetry observations such that $\mathbf{s}(t_i)\sim \text{N}(\boldsymbol\mu(t_i),\sigma^2\mathbf{I})$ (for $i=1,\ldots,n$) and a mixture SDE with components 

\begin{equation}           
  d\boldsymbol\mu(t) = 
    \begin{cases}
      \sigma_0 d\mathbf{b}_0(t) &\mbox{, } z(t)=0 \\
      -\bigtriangledown p(\boldsymbol\mu(t),\boldsymbol\beta)dt + \sigma_1 d\mathbf{b}_1(t) &\mbox{, } z(t)=1 
    \end{cases} \,, \label{eq:sde_mix} 
\end{equation}           
\noindent for the set of times in the study period $t\in {\cal T}$, where, $p$ represents a potential function (so-called because of its connections to potential energy in physics; Preisler et al., 2013) controlling the drift of the individual trajectory $\boldsymbol\mu(t)$ based on landscape covariates and associated coefficients $\boldsymbol\beta$.  The diffusion aspects of the movement process are controlled by the two Gaussian white noise terms $d\mathbf{b}_0(t)$ and $d\mathbf{b}_1(t)$ that are scaled by $\sigma_0$ and $\sigma_1$.  

\begin{figure}[htp]
\centering
\includegraphics[width=6in]{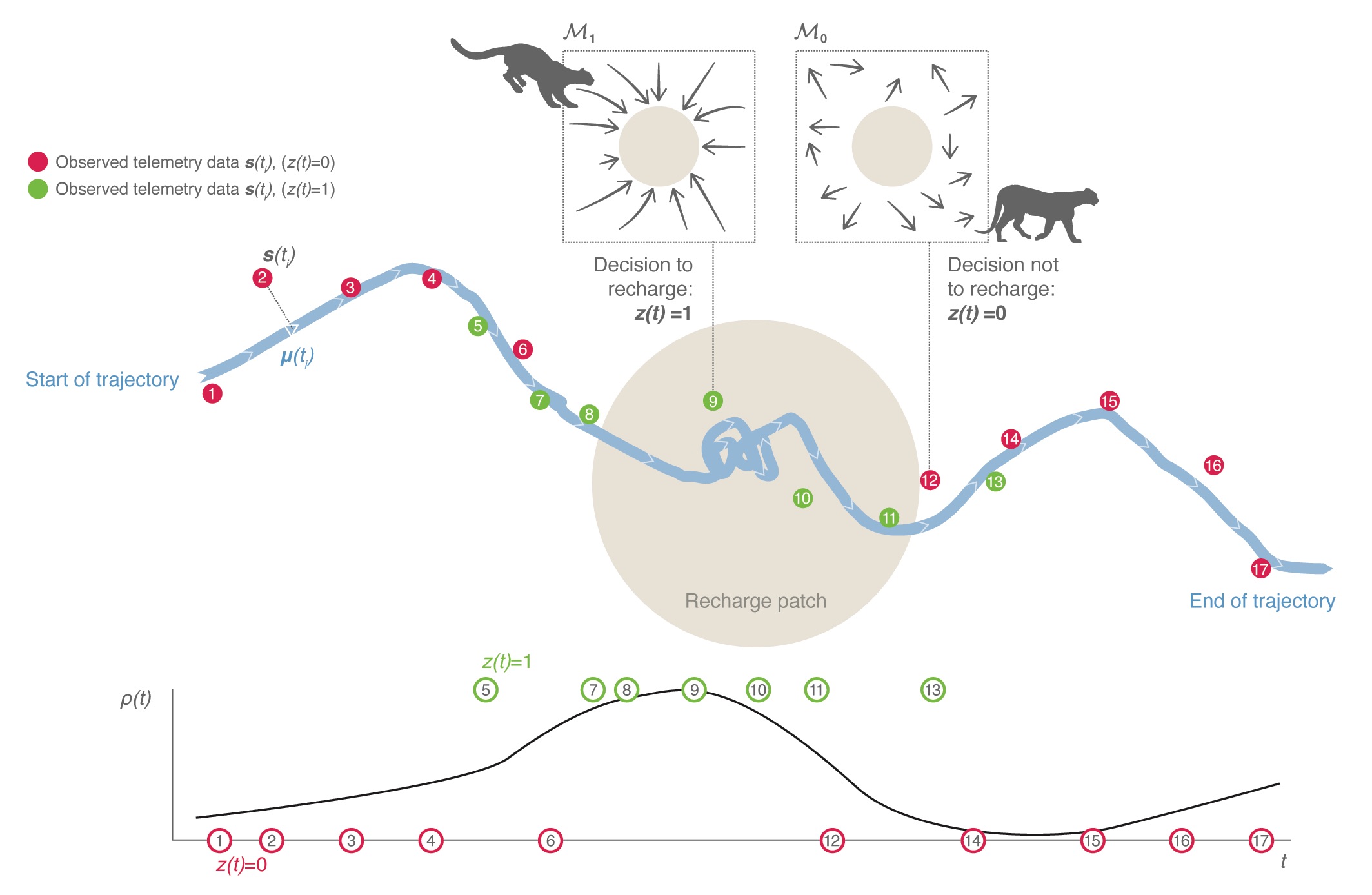}
\caption{\scriptsize Schematic of recharge and movement model components.  The observed telemetry data ($\mathbf{s}(t_i)$, red and green points along trajectory) at time $t_i$ are measurements (with error) of the true positions $\boldsymbol\mu(t_i)$ (blue triangle, left, for a given observation time $t_i$).  The underlying continuous-time trajectory $\boldsymbol\mu(t)$ is shown as the solid blue line and is conditionally modeled based on the movement dynamics (incorporated in models ${\cal M}_0$ and ${\cal M}_1$) and possibly changes in the environment (incorporated in model ${\cal M}_1$). In this example, the brown circle in the middle of the study area represents a recharge region or patch where the individual may recharge its energy (e.g., a prey kill area).  The binary decision $z(t)$ to recharge indicates when the individual responds to the underlying landscape (in this case, it may be attracted to the recharge region).  While $z(t)$ is represented as a continuous-time binary process in our model, this figure shows the subset of decisions associated with the observed telemetry data (numbered points in bottom plot).  In the figure, decisions to recharge ($z(t)=1$) are green and are otherwise shown in red ($z(t)=0$).  The stochastic binary decision process is governed by the probability function $\rho(t)$ (shown as solid black line in bottom plot), which is, in turn, a function of the recharge process $g(t)$ (not shown).}
\label{fig:model}
\end{figure}

The movement process model in (\ref{eq:sde_mix}) can be interpreted in the following way.  When the decision to recharge is made ($z(t)=1$) at time $t$, the individual will respond to the environment as dictated by the potential function $p(\boldsymbol\mu(t),\boldsymbol\beta)$ by taking steps that are aligned approximately with its gradient surface (i.e., downhill on the surface; ${\cal M}_1$ in Figure~\ref{fig:model}).  Conversely, when $z(t)=0$, the individual may roam freely without needing to respond to the environment (${\cal M}_0$ in Figure~\ref{fig:model}).  Thus, in this particular model specification, we would obtain biased inference about the movement parameters $\boldsymbol\beta$ if the individual was assumed to move according to the SDE with potential function $p(\boldsymbol\mu(t),\boldsymbol\beta)$ without considering the underlying physiological process (i.e., $z(t)=1$ always).  Most studies investigating resource selection assume only a single movement model.  Thus, the movement model specification in (\ref{eq:sde_mix}) allows us to infer when a physiological signal is present in our telemetry data (i.e., when $z(t)$ switches between zero and one at some point along the trajectory).  

It is worth noting that our model formulation fits into a broader class of models for movement using the basis function approach proposed by Hooten and Johnson (2017a) to connect the telemetry data to the underlying trajectory.  This framework provides opportunities to extend the model in future studies to accommodate other types of smoothness and heterogeneity in the trajectory process (see Scharf et al., 2018 and Hooten et al., 2018 for further details).  Also, to fit the model to data, we must solve the SDE for $\boldsymbol\mu(t)$ based on a discrete approximation.  This solution is more intuitive than the SDE itself because it assumes a discrete-time form where the process components of the model for $\boldsymbol\mu(t_j)$ in (\ref{eq:process_model}) can be written as ${\cal M}_0=\text{N}(\boldsymbol\mu(t_{j-1}),\sigma^2_0\mathbf{I}\Delta t)$ and ${\cal M}_1=\text{N}(\boldsymbol\mu(t_{j-1})-\bigtriangledown p(\mathbf{x}'(\boldsymbol\mu(t_{j-1}))\boldsymbol\beta)\Delta t,\sigma^2_1\mathbf{I}\Delta t)$ for a fine grid of time points, $t_1,\ldots,t_m$, spaced $\Delta t$ apart, using an Euler-Maruyama discretization scheme (Kloeden and Platen, 1992).  

As a result of our specifications for the hierarchical model, the full parameter set includes the latent position process $\boldsymbol\mu(t_j)$ for all $j=1,\ldots,m$, as well as 3 sets of parameters:  1) the drift coefficients in the potential function, $\boldsymbol\beta$, 2) the initial recharge state $g_0$ and recharge coefficients $\boldsymbol\theta$, and 3) the variance parameters $\sigma^2_s$, $\sigma^2_0$, and $\sigma^2_1$.  To estimate the parameters and make inference, we can fit the model using maximum likelihood if we are able to derive the integrated likelihood, or we can use Bayesian methods.  In what follows, we use a Bayesian approach that allows us to specify priors for the three sets of parameters described above (Online Appendix C) and obtain a Markov chain Monte Carlo (MCMC) sample from the posterior distribution  

\begin{align}
  [\{\boldsymbol\mu(t_j), &\text{ for } j=1,\ldots,m\}, \boldsymbol\beta, g_0, \boldsymbol\theta, \sigma^2_s, \sigma^2_0, \sigma^2_1 | \{\mathbf{s}(t_i), i=1,\ldots,n\}]\propto \notag \\
  &\prod_{i=1}^n [\mathbf{s}(t_i) | \boldsymbol\mu(t_i),\sigma^2_s] \prod_{j=1}^m [\boldsymbol\mu(t_j)|\boldsymbol\mu(t_{j-1}),\sigma^2_0]^{1-z(t_j)}[\boldsymbol\mu(t_j)|\boldsymbol\mu(t_{j-1}),\boldsymbol\beta,\sigma^2_1]^{z(t_j)}[z(t_j)|g_0,\boldsymbol\theta] \times \notag \\
  &[\boldsymbol\beta][g_0][\boldsymbol\theta][\sigma^2_s][\sigma^2_0][\sigma^2_1] \,, \label{eq:post}
\end{align}
\noindent for a fine discretization of the latent position process $\boldsymbol\mu(t_j)$ at times $t_1,\ldots,t_m$ and where we have suppressed notation for the position process in the conditional distribution for $z(t_j)$ to streamline the expression.  

We applied specific formulations of our hierarchical movement model to infer recharge dynamics based on telemetry data for two contrasting species: a mountain lion in the Front Range of the Rocky Mountains in Colorado, USA and an African buffalo in Kruger National Park, South Africa (Figure~\ref{fig:map}).  Also, for illustration, we demonstrate the approach based on simulated data in Online Appendix B.  Using simulated data, we showed that the modeling framework allows us to recover parameters and identify the data generating model compared to a set of alternatives that consider only ${\cal M}_0$ and ${\cal M}_1$ individually (Online Appendix B).    

\section{Results}
\subsection{Mountain Lion}
\begin{figure}[htp]
\centering
\includegraphics[width=6in]{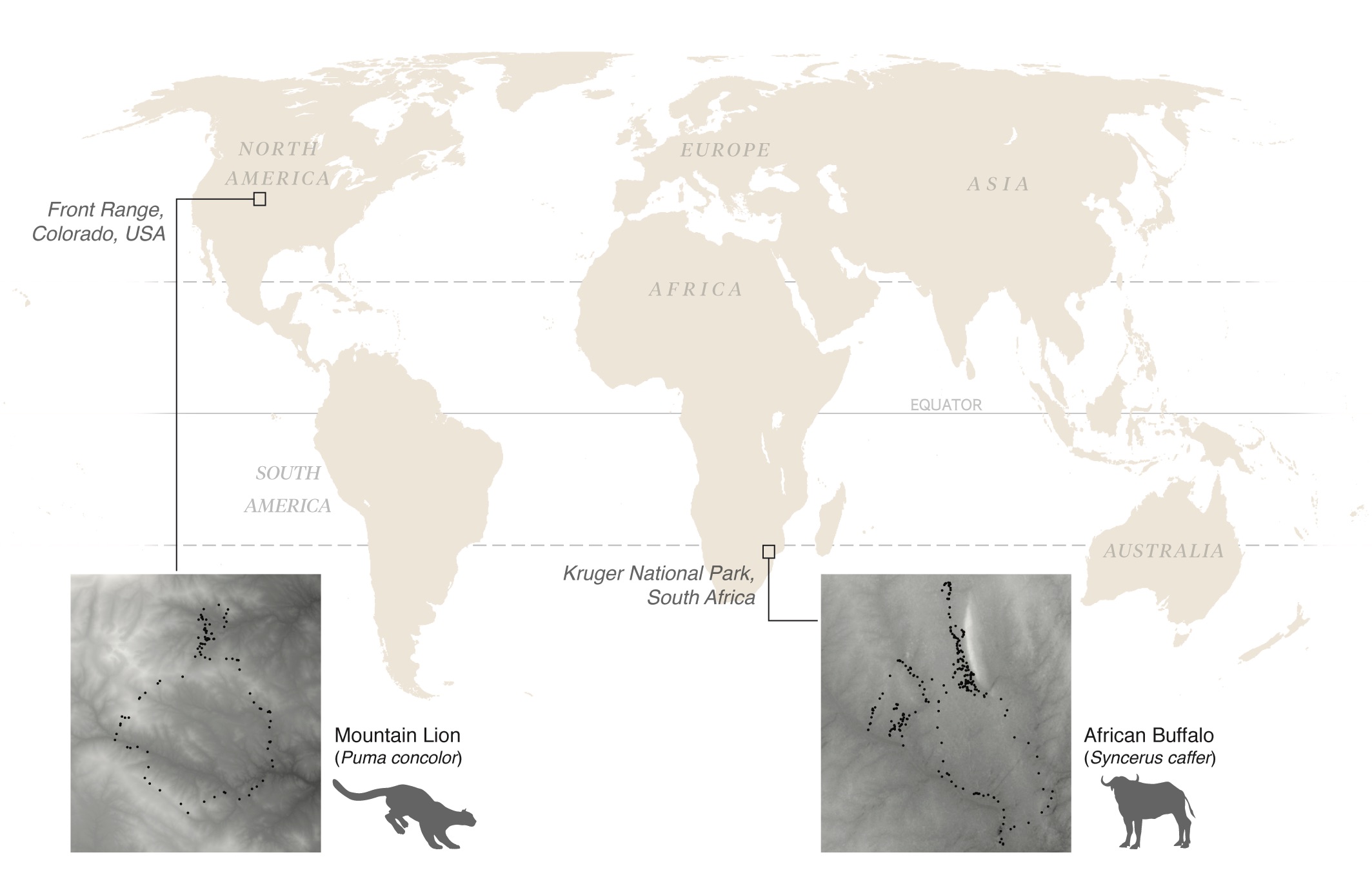}
\caption{World map depicting the regions where the telemetry data in our examples arise from a GPS collared mountain lion and African buffalo.  Telemetry data are shown as black points on blown up maps, with elevation shown as background shading; high (relative) elevations shown as lighter shading.}
\label{fig:map}
\end{figure}

In the western USA, mountain lions (\emph{Puma concolor}) are apex predators that mostly seek mule deer (\emph{Odocoileus hemionus}) as prey.  In the Front Range of the Rocky Mountains in Colorado, USA (Figure~\ref{fig:map}), many approaches have been used to model the individual-based movement of mountain lions (e.g., Hanks et al., 2015; Hooten and Johnson, 2017a; Buderman et al., 2018), but none have modeled connections between physiological dynamics and movement.  Front Range mountain lions navigate a matrix of public and privately owned land comprised of wildland-urban interface, roads, and trail systems (Blecha, 2015; Buderman et al., 2018).  Previous research has shown that prey availability and cached carcasses are important factors influencing mountain lion movement (Husseman et al., 2003; Blake and Gese, 2016).  Thus, we specified a recharge-based movement model for the telemetry data (global positioning system [GPS] with 3 hr fixes; n=150) from an adult male mountain lion in Colorado during April 25, 2011 -- May 17, 2011 (Figure~\ref{fig:map}).  

This particular trajectory includes a period at the beginning and end of the time interval where the individual occupied a prey kill area (top center of blown up region in Figure~\ref{fig:map}; using methods to identify kills sites described by Knopff et al., 2009).  On approximately May 1, 2011, the individual mountain lion left the prey kill area to traverse a large loop to the south before returning to the prey kill area.  After a few more days at the prey kill area, the individual left again to traverse a small loop to the north.  We hypothesized that the mountain lion individual recharged at the prey kill area and mostly discharged otherwise. 

We used the same movement model structure as specified in the previous section, with ${\cal M}_0$ implying no drift when charged and $p(\boldsymbol\mu(t),\boldsymbol\beta)=\mathbf{x}'(\boldsymbol\mu(t))\boldsymbol\beta$ to account for drift when discharged.  To formulate the recharge component of the full model, we used an intercept ($\theta_0$), and six spatial covariates:  presence in the prey kill area, elevation, slope, sine and cosine of aspect, and the interaction of elevation and slope.  For movement covariates in the full model, we used five:  elevation, slope, sine and cosine of aspect, and distance to prey kill area.  

\begin{figure}[htp]
\centering
\includegraphics[width=6in]{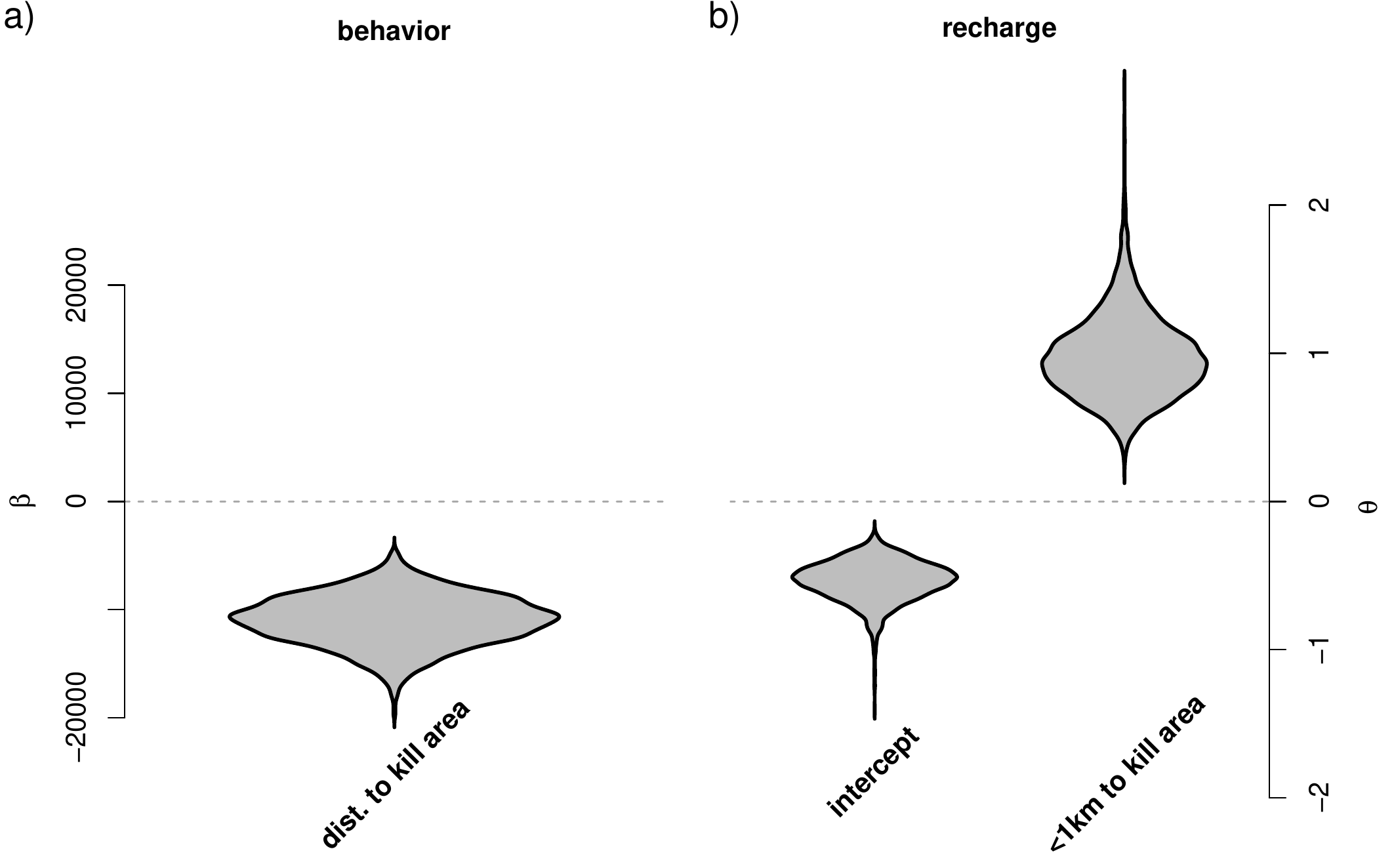}
\caption{Marginal posterior violin plots for the mountain lion model parameters a) $\boldsymbol\beta$ and b) $\boldsymbol\theta$.}
\label{fig:ml_post_params}
\end{figure}

We fit the full recharge-based movement model to the mountain lion telemetry data shown in Figure~\ref{fig:map}.  The set of priors and hyperparameter settings, as well as pseudocode and computational details to fit the recharge-based movement model, are provided in Online Appendix C.  We also examined a set of simpler models including model ${\cal M}_0$ and ${\cal M}_1$ separately as well as the recharge-based model with only prey kill area covariates and the associated submodel ${\cal M}_1$ with only the prey kill area covariates.  We scored each of the models using the negative log posterior predictive score based on cross-validation (Online Appendix C) and found the recharge-based model with only prey kill area covariates was the best predictive model.  The associated marginal posterior distributions for the model parameters $\beta_1$ (coefficient for distance to prey kill area), $\theta_0$ (recharge intercept coefficient), and $\theta_1$ (coefficient for inside prey kill area) are shown in Figure~\ref{fig:ml_post_params}.  

In this case, the left half of Figure~\ref{fig:ml_post_params} (labeled ``behavior'') indicates that there is evidence for the individual to move toward the prey kill area when the decision to recharge is made (because of the negative coefficient associated with distance to prey kill area) and the recharge function itself (and hence the decision to recharge) increased with the individual's presence in the prey kill area (i.e., convex polygon with 1 km buffer from kill site clusters).      

\begin{figure}[htp]
\centering
\includegraphics[width=5in]{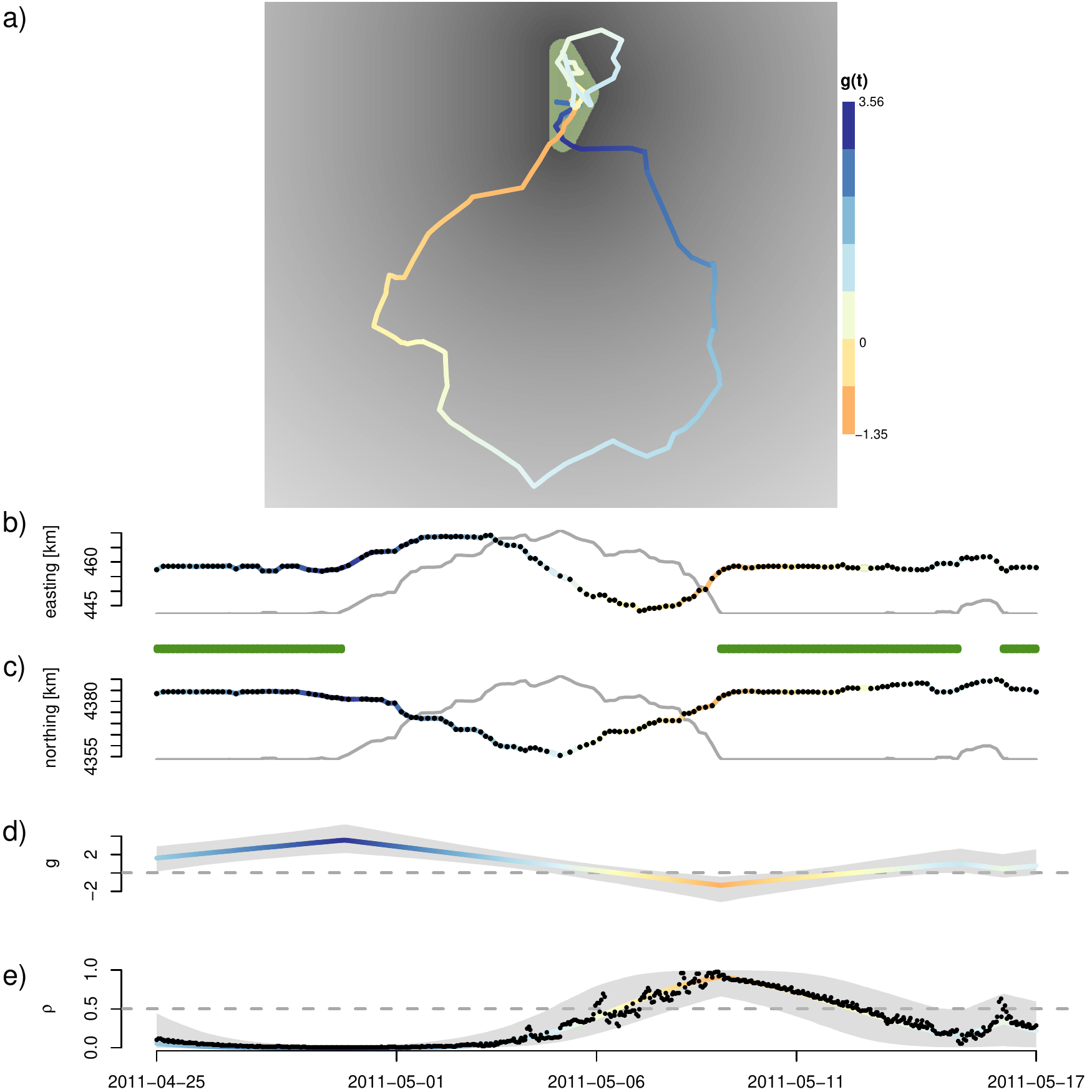}
\caption{\scriptsize Posterior median associated with the mountain lion data analysis for the a) recharge function $g(t)$ shown as color on top of the posterior mean trajectory $\boldsymbol\mu(t)$.  Prey kill area (i.e., convex polygon with 1 km buffer from prey kill site clusters) shown as green region indicating area associated with recharge.  Distance to prey kill area is shown in the background for reference (with small distances indicated by darker shades).  Map in (a) oriented such that north is up.  Posterior median trajectories (b, c) and d) recharge function $g(t)$ and e) decision probability $\rho(t)$ with 95\% credible intervals shown in gray with posterior mean for the decision $z(t)$ shown as black points.  Color corresponds to the value of the recharge function.  Profile of distance to prey kill area shown as gray line in (b) and (c) for reference.  Green rug at the bottom of (b) represents times when recharge occurred.}
\label{fig:ml_post_grho}
\end{figure}

In terms of the estimated recharge function for the individual mountain lion, the posterior median for $g(t)$ is shown superimposed on the trajectory in Figure~\ref{fig:ml_post_grho}.  The results of fitting the recharge-based movement model to the mountain lion telemetry data indicate that the individual is charged (blue) when near the prey kill area (green region) and discharges as it moves farther from the kill area, both to the south and the north (Figure~\ref{fig:ml_post_grho}a).

Visualized longitudinally, the posterior marginal trajectories as well as posterior median for $g(t)$ and $\rho(t)$ are shown in Figure~\ref{fig:ml_post_grho}. The posterior inference indicates that the mountain lion individual we analyzed was mostly recharging during the early portion of the study period (April 25, 2011 --- May 1, 2011).  However, as the recharge function $g(t)$ exceeded a value of approximately three, the individual left the prey kill area.  During the week that the individual was away from the prey kill area, our analysis shows that the aggregated physiological process discharged until the behavioral decision process was dominated by $z(t)=1$, at which point the individual actively sought to recharge.  This decision process was characterized largely by a tendency of the individual to orient back toward the prey kill area on May 9, 2011 (Figure~\ref{fig:ml_post_grho}).  Then, after another few days of recharging at the original prey kill area, the individual left the prey kill area again (this time to the north) and its physiological process began to discharge again until near the end of the study period when the individual returned to the prey kill area (Figure~\ref{fig:ml_post_grho}).  

\subsection{African Buffalo}
In contrast to the western hemisphere predator we described in the previous section, the African buffalo is a large grazing ungulate that ranges throughout sub-Saharan Africa (Sinclair, 1977).  In Kruger National Park, South Africa, the African buffalo is an important species because it fills a niche in terms of tall and coarse grazing preference (Corn\'elis et al., 2014), is a source of prey for lions (\emph{Panthera leo}; Sinclair, 1977; Prins, 1996; Radloff and DuToit, 2004), and is one of the desirable species for tourism in the region.  African buffalo are strongly water dependent because they lack the capacity to subsist on the moisture available from their forage alone (Prins and Sinclair, 2013).  Previous studies of the movement of African buffalo found that water resources can strongly influence their space use (Redfern et al., 2003).  In some cases, African buffalo may undergo large interseasonal movements when resources are limited (e.g., Naidoo et al., 2012), but there is variability in dry versus wet season movement characteristics across regions (Ryan et al., 2006; Corn\'elis et al., 2014).  Repetitive use of areas is common among African buffalo and some of these patterns in space use may be a result of maintaining physiological balance among resources (Bar-David et al., 2009). 

We used the same movement model (\ref{eq:sde_mix}) that we applied to the mountain lion data (but with different environmental variables) to analyze a set of telemetry data arising from an adult female African buffalo in southern Kruger National Park (Getz et al., 2007) obtained using hourly GPS fixes (n=361) and spanning the period from October 1, 2005 -- October 14, 2005 (Figure~\ref{fig:map}).  The transition from dry to wet season typically occurs during late September and October in South Africa, and the year 2005 had slightly more rainfall than the climate average for Kruger National Park (MacFadyen et al., 2018).  The African buffalo movement data we analyzed indicates that the individual mostly occupied the northern and western extent of the region during the two week time period, but traveled approximately 15 km between major surface water sources to the southeastern portion of the region during October 7--9, 2005.  

We specified the recharge function to include an intercept ($\theta_0$) and covariates for elevation, slope, surface water proximity ($<0.5$ km buffer to nearest surface water), and an interaction for elevation$\times$slope to examine the evidence for an effect of water and other resources for which topography may serve as a surrogate on physiological recharge during a time when it is difficult to predict the widespread availability of water and forage during the transition from dry to wet season in this region.  For movement covariates, we used elevation, slope, and distance to nearest surface water.         

We fit recharge-based hierarchical movement models to the African buffalo telemetry data shown in Figure~\ref{fig:map}.  The full set of priors and hyperparameter settings, as well as pseudocode and computational details to fit the full recharge-based movement model, are provided in Online Appendix C.  As in the mountain lion data analysis, we also examined a set of simpler models, including hierarchical models that incorporate ${\cal M}_0$ and ${\cal M}_1$ separately with all covariates as well as ${\cal M}_1$ with only surface water covariates both together with ${\cal M}_0$ and separately.   

\begin{figure}[htp]
\centering
\includegraphics[width=6in]{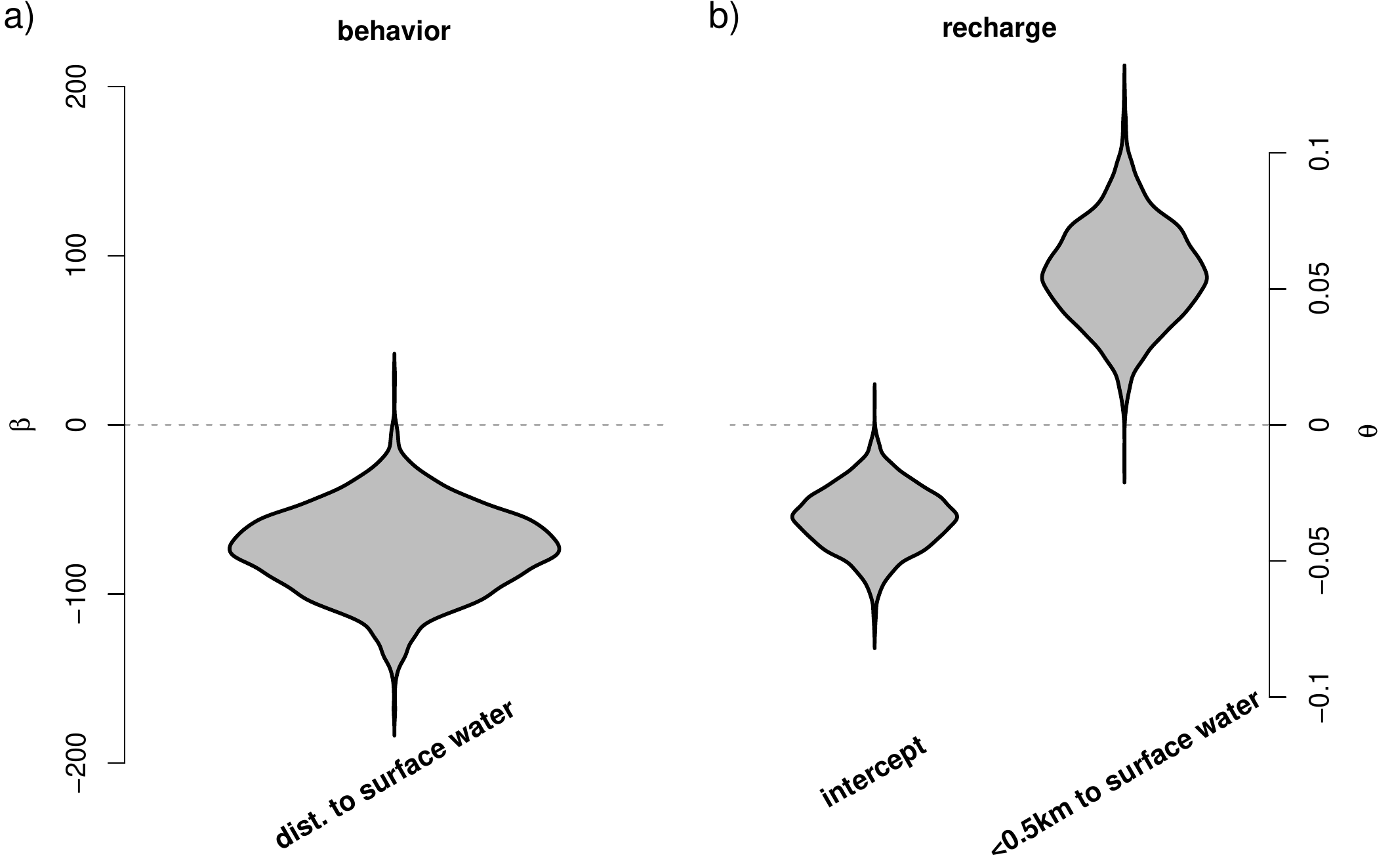}
\caption{Marginal posterior violin plots for the African buffalo model parameters a) $\boldsymbol\beta$ and b) $\boldsymbol\theta$.}
\label{fig:ab_post_params}
\end{figure}

Similar to our mountain lion results, the reduced recharge model based only on surface water covariates had a better predictive score than the other models we fit (Online Appendix C).  The left half of Figure~\ref{fig:ab_post_params} (labeled ``behavior''), which shows the marginal posterior distribution for the movement parameter, indicates that the African buffalo orients toward surface water when it makes the decision to recharge during this time period.  Furthermore, the right half of Figure~\ref{fig:ab_post_params} indicates that surface water proximity increased the recharge function itself.  These results agree with previous findings (e.g., Redfern et al., 2003) that surface water in this region is an important predictor of African buffalo movement.  

\begin{figure}[htp]
\centering
\includegraphics[width=6in]{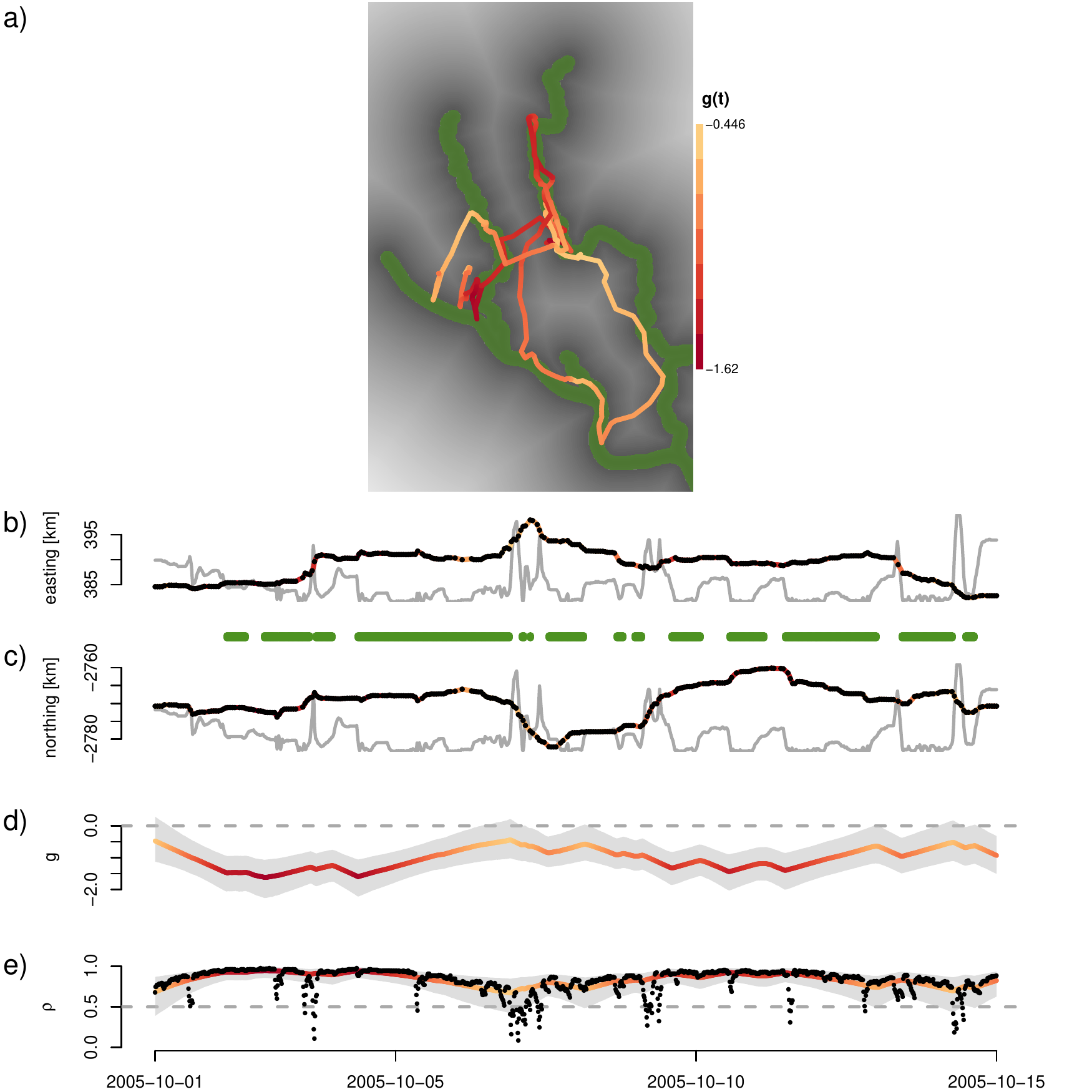}
\caption{\scriptsize Posterior median associated with the African buffalo data analysis for the a) recharge function $g(t)$ shown as color on top of the posterior mean trajectory $\boldsymbol\mu(t)$.  Distance to surface water is shown in the background for reference (with small distances indicated by darker shades) and green indicating inferred areas associated with recharge.  Map in (a) oriented such that north is up.  Posterior median trajectories (b, c) and d) recharge function $g(t)$ and e) decision probability $\rho(t)$ with 95\% credible intervals shown in gray with posterior mean for the decision $z(t)$ shown as black points.  Color corresponds to the value of the recharge function.  Distance to surface water profile shown as gray line in (b) and (c) for reference.  Green rug at the bottom of (b) represents times when the recharge function is increasing.}
\label{fig:ab_post_grho}
\end{figure}

Displayed in the same way as the mountain lion results, Figure~\ref{fig:ab_post_grho} shows the posterior marginal trajectories as well as posterior median for $g(t)$ and $\rho(t)$ for the African buffalo.  The posterior inference indicates that the African buffalo individual we analyzed needed to recharge regularly throughout the time period based on the large values for $\rho(t)$ overall.  However, brief and fairly regular periods where the posterior mean for $z(t)$ dropped below $0.5$ in Figure~\ref{fig:ab_post_grho}e indicate short forays away from water resources.  One such period where the decision process was not dominated by $z(t)=1$ occurred when the individual looped to the southeast of the study area (October 7--8, 2005).  Our analysis shows that the recharge function started high (near zero) and then mostly decreased as the individual ventured farther from surface water until eventually looping back to the north at which point the recharge function increased again (Figure~\ref{fig:ab_post_grho}a,d).  In fact, Figure~\ref{fig:ab_post_grho}a shows the areas associated with increases in the recharge function in green.  This spatially-explicit inference indicates that low lying areas near the Sabie River and tributaries are associated with recharge for the African buffalo individual we analyzed (Figure~\ref{fig:ab_post_grho}a).  Furthermore, the fact that the recharge model including surface water proximity covariates had a better predictive score than the simpler models (${\cal M}_0$ and ${\cal M}_1$) fit separately, suggests that a physiological recharge signal related to the covariates is present in the movement trajectory for the African buffalo.     

\section{Discussion}
Our example data analyses provided evidence that both the mountain lion and African buffalo data sets contained a physiological signal whose variation is at least partially explained by environmental features.  In the case of the mountain lion, a model comparison indicated that proximity to prey kill area was the primary factor influencing the recharge and movement processes.  This result agrees with other recent studies (i.e., Buderman et al., 2018) that mountain lion movement patterns are strongly influenced by predatory behavior.  Our analysis of the African buffalo data suggested that recharge-based dynamics were important because the simpler models that do not directly account for an underlying physiological process had worse predictive scores.  In the case of the African buffalo data we analyzed, the inferred spatial pattern associated with recharge in Figure~\ref{fig:ab_post_grho}a indicated a clear relationship between probable surface water and recharge and this was confirmed by the posterior distributions for movement and recharge parameters (Figure~\ref{fig:ab_post_params}).  Previous studies of African buffalo indicate that, while movement is largely driven by water resources, other factors such as forage, social dynamics, and cover may also influence space use (Ryan et al., 2006; Winnie et al., 2008).  These additional factors could be examined in more detailed studies that combine recharge and social dynamics with plant ecology and energetics.   

In general, the feedback between animal decision making, physiology, and movement is a complex process that involves both intrinsic and extrinsic factors (Morales et al., 2005; Nathan at al., 2008; Morales et al., 2010).  For example, connections between energetics, memory, and movement directly influence the way we infer animal home ranges (B\"orger et al., 2008).  Despite calls for more thoughtful frameworks to model movement that consider mechanisms explicitly, many modern approaches to modeling animal trajectories are still purely phenomenological.  Recent advances in biotelemetry technology have given rise to massive repositories of high-resolution individual-based data (``auxiliary data'') that often accompany more conventional position-based telemetry data (Brown et al., 2013).  These auxiliary data are collected to measure characteristics of individual fitness and behavior (e.g., Elliot et al., 2013; Leos-Barajas et al., 2017) and may provide a more direct link to understand physiological recharge.     

Leveraging the hierarchical modeling framework to combine data sources (Hobbs and Hooten, 2015), we can integrate auxiliary data into the recharge-based animal movement model (Online Appendix E).  Such model structures have become common in population and community ecology where they are referred to as ``integrated population models'' (Schaub and Abadi, 2011).  When we have auxiliary accelerometer data, it may be possible to connect the fine-scale measurements of micro-movement to the change in position directly (Wilson et al., 1991).  In that case, it is sensible to let the auxiliary data inform both the trajectory process and the physiological recharge process directly.  In situations where multiple forms of auxiliary data are recorded (e.g., accelerometer and body condition measurements), we can augment the integrated movement model with additional data models that are connected to the latent model components, partitioning the recharge functions further as needed (Online Appendix E).

Overall, the framework we present allows researchers to connect the mechanisms related to known physiological characteristics with more conventional telemetry data to account for latent physiological and individual-based decision processes.  Our approach is flexible and allows for modifications to the form of both movement (\ref{eq:data_model})--(\ref{eq:process_model}) and recharge functions (\ref{eq:gvt}) and (\ref{eq:gt}).  As with any mixture model, some structure allows the data to better separate model components so that parameters are identifiable.  In our case studies, we specified the movement model such that one term (${\cal M}_0$) represents random diffusive movement and the other term (${\cal M}_1$) captures movement in response to environmental variables.  This helps us learn about the recharge function in a way that corresponds to our preexisting knowledge about the physiology of these species.  In Online Appendix E, we show how to extend the recharge-based movement model to accommodate various sources of auxiliary data to better recognize and estimate the physiological process components depending on available data.     

For some species, it may be appropriate to consider additional stochasticity in the recharge process because of unobservable interactions with conspecifics, allospecifics, or other dynamic environmental conditions.  Our framework can readily accommodate these sources of overdispersion by specifying the recharge functions $g(v,t)$ as SDEs (in addition to the movement process).  Statistical inference in these settings relies on our ability to observe enough data to successfully estimate the various sources of uncertainty in the model.  Auxiliary data, such as those described above, may be helpful to partition and estimate parameters in these more general models.

We formulated the recharge-based movement models in continuous time for our applications to account for irregular telemetry and auxiliary data when available, but, like all continuous-time models that require numerical solutions, our model is fit using an intuitive discrete time approximation.  In cases where the telemetry data are high-resolution and temporally regular, the movement models (i.e., ${\cal M}_0$ and ${\cal M}_1$) themselves can be formulated directly in discrete time using either the velocity vectors (e.g., Jonsen et al., 2005) or polar coordinates associated with discrete moves (e.g., Morales et al., 2004; Langrock et al., 2012; McClintock et al., 2012).  In this setting, the movement process and physiological recharge function are limited to the chosen temporal resolution and the associated inference is resolution-dependent.    

While our recharge-based movement modeling framework facilitates the inclusion of mechanisms related to physiology, it can also be used as a way to accommodate latent sources of dependence.  The physiological recharge functions we specified in (\ref{eq:gvt}) and (\ref{eq:gt}) impart a type of long memory in the stochastic process models that we exploit to learn about the influences of landscape and other spatial features on movement.  However, time series analyses have relied on long-memory processes to account for dependence in data for many other applications (Beran, 1994).  In terms of animal memory explicitly, its influence on movement has been investigated separately (e.g., Fagan et al., 2013; Avgar et al., 2013; Bracis et al., 2015; Bracis and Mueller, 2017; Merkle et al., 2017), but it has not been accommodated in the way we describe herein, especially in the context of physiological processes.       

\section*{Acknowledgements}
The authors thank the editors and three anonymous reviewers whose comments helped improve this work.  The authors also thank Jake Ivan, Mat Alldredge, Ephraim Hanks, Franny Buderman, Devin Johnson, Daisy Chung, and Brett McClintock for numerous helpful discussions and previous research in this area.  This research was funded by NSF DMS 1614392 (MBH) and PICT 2015 0815 (JMM).  Any use of trade, firm, or product names is for descriptive purposes only and does not imply endorsement by the U.S. Government.  Data and computer code are available at:  

\verb!https://github.com/henryrscharf/Hooten_et_al_EL_2018!. 

\section*{References}

\rf Alcock, J. 2009. Animal Behavior: An Evolutionary Approach. Sinauer.  Sunderland, Massachusetts, USA.

\rf Alexander, R.M. 2003.  Principles of Animal Locomotion.  Princeton University Press.  Princeton, New Jersey, USA.

\rf Avgar, T., R. Deardon, and J.M. Fryxell.  (2013).  An empirically parameterized individual based model of animal movement, perception, and memory.  Ecological Modeling, 251: 158-172. 

\rf Bar-David, S., I. Bar-David, P.C. Cross, S.J. Ryan, C.U. Knechtel, and W.M. Getz. (2009). Methods for assessing movement path recursion with application to African buffalo in South Africa. Ecology, 90: 2467-2479.

\rf Bracis, C., E. Gurarie, B. Van Moorter, and R.A. Goodwin. (2015). Memory effects on movement behavior in animal foraging. PloS One, 10: e0136057. 

\rf Bracis, C. and T. Mueller.  (2017). Memory, not just perception, plays an important role in terrestrial mammalian migration. Proceedings of the Royal Society B, 284: 20170449. 

\rf Beran, J.  (1994).  Statistics for Long-Memory Processes.  Chapman \& Hall/CRC.

\rf Blake, L.W. and E.M. Gese. (2016). Resource selection by cougars: Influence of behavioral state and season. The Journal of Wildlife Management, 80: 1205-1217.

\rf Blecha, K.A. (2015). Risk-reward tradeoffs in the foraging strategy of cougar (Puma concolor): Prey distribution, anthropogenic development, and patch selection. Thesis, Colorado State University, Fort Collins, CO.

\rf Bleicher, S.S. (2017).  The landscape of fear conceptual framework:  Definition and review of current applications and misuses.  PeerJ, 5: e3772.

\rf B\"orger, L., B.D. Dalziel, and J.M. Fryxell. (2008). Are there general mechanisms of animal home range behaviour? A review and prospects for future research. Ecology Letters, 11: 637-650.

\rf Brown, D., R. Kays, M. Wikelski, R. Wilson, R., and A.P. Klimley. (2013). Observing the unwatchable through acceleration logging of animal behavior. Animal Biotelemetry, 1: 1-16.

\rf Buderman, F.E., M.B. Hooten, M. Aldredge, and J.S. Ivan. (2018). Time-varying predatory behavior is primary predictor of fine-scale movement of wildland-urban cougars.  Movement Ecology, 6: 22. 

\rf Brillinger, D.R. (2010). Modeling spatial trajectories. In Gelfand, A. E., P. J. Diggle, M. Fuentes, and P. Guttorp, editors, Handbook of Spatial Statistics, chapter 26, pages 463—-475. Chapman \& Hall/CRC, Boca Raton, Florida, USA.  

\rf Cagnacci, F., L. Boitani, R.A. Powell, and M.S. Boyce.  (2010).  Animal ecology meetings GPS-based radiotelemetry:  A perfect storm of opportunities and challenges.  Philosophical Transactions of the Royal Society of London B: Biological Sciences, 365: 2157-2162.

\rf Colgan, P. (1989).  Animal Motivation.  Springer Netherlands.

\rf Cooke, S.J., S.G. Hinch, M. Wikelski, R.D. Andrews, L.J. Kuchel, T.G. Wolcott, and P.J. Butler. (2004). Biotelemetry: a mechanistic approach to ecology. Trends in Ecology and Evolution, 19: 334-343.

\rf Cornelis, D., M. Melletti, L. Korte, S.J. Ryan, M. Mirabile, T. Prin, and H.H.T. Prins. (2014). African Buffalo \emph{Syncerus caffer}. In: Ecology, Evolution and Behaviour of Wild Cattle.  Melletti, M. and J. Burton (eds.).  Cambridge University Press.  

\rf Dudeck, B.P., M. Clinchy, M.C. Allen, and L.Y. Zanette. (2018). Fear affects parental care, which predicts juvenile survival and exacerbates the total cost of fear on demography. Ecology, 99: 127-135.

\rf Eggleston, D.B. and R.N. Lipcius. (1992). Shelter selection by spiny lobster under variable predation risk, social conditions, and shelter size. Ecology, 73: 992-1011.

\rf Elliott, K.H., M. Le Vaillant, A. Kato, J.R. Speakman, and Y. Ropert-Coudert. (2013). Accelerometry predicts daily energy expenditure in a bird with high activity levels. Biology Letters, 9: 20120919.

\rf Fagan, W.F., M.A. Lewis, M. Auger‐M\'eth\'e, T. Avgar, S. Benhamou, G. Breed, L. LaDage, U.E. Schlägel, W.W. Tang, Y.P. Papastamatiou,  and J. Forester.  (2013). Spatial memory and animal movement. Ecology Letters, 16: 1316-1329.

\rf Full, R., D. Zuccarello, and A. Tullis. (1990). Effect of variation in form on the cost of terrestrial locomotion.  Journal of Experimental Biology, 150: 233-246.

\rf Gelfand, A.E. and A.F. Smith. (1990). Sampling-based approaches to calculating marginal densities. Journal of the American Statistical Association, 85: 398-409.

\rf Getz, W.M., S. Fortmann-Roe, P.C. Cross, A.J. Lyons, S.J. Ryan, and C.C. Wilmers. (2007). LoCoH: nonparameteric kernel methods for constructing home ranges and utilization distributions. PloS one, 2(2), e207.

\rf Green, J.A. (2011). The heart rate method for estimating metabolic rate: review and recommendations. Comparative Biochemistry and Physiology Part A: Molecular and Integrative Physiology, 158: 287-304.

\rf Halsey, L.G. (2016).  Terrestrial movement energetics:  Current knowledge and its application to the optimising animal.  Journal of Experimental Biology, 219: 1424-1431. 

\rf Hanks, E.M., M.B. Hooten, and M. Alldredge. (2015). Continuous-time discrete-space models for animal movement. Annals of Applied Statistics, 9: 145-165. 

\rf Hobbs, N.T. and M.B. Hooten. (2015). Bayesian Models: A Statistical Primer for Ecologists. Princeton University Press.

\rf Hooten, M.B. and N.T. Hobbs. (2015). A guide to Bayesian model selection for ecologists. Ecological Monographs, 85: 3-28.

\rf Hooten, M.B. and D.S. Johnson. (2017a). Basis function models for animal movement. Journal of the American Statistical Association, 112: 578-589.

\rf Hooten, M.B. and D.S. Johnson. (2019). Modeling Animal Movement. Gelfand, A.E., M. Fuentes, and J.A. Hoeting (eds). In Handbook of Environmental and Ecological Statistics. Chapman and Hall/CRC. 

\rf Hooten, M.B., D.S. Johnson, B.T. McClintock, and J.M. Morales. (2017). Animal Movement: Statistical Models for Telemetry Data. Chapman and Hall/CRC.

\rf Hooten, M.B., H.R. Scharf, T.J. Hefley, A. Pearse, and M. Weegman. (2018). Animal movement models for migratory individuals and groups. Methods in Ecology and Evolution, 9: 1692-1705. 

\rf Houston A.I. and J.M. McNamara. (1999). Models of Adaptive Behaviour. Cambridge University Press.  Cambridge, United Kingdom.

\rf Humphries, M.M. and V. Careau. (2011). Heat for nothing or activity for free? Evidence and implications of activity-thermoregulatory heat substitution. Integrative and Comparative Biology, 51: 419-431.

\rf Husseman, J.S., D.L. Murray, G. Power, C. Mack, C. Wenger, and H. Quigley. (2003). Assessing differential prey selection patterns between two sympatric large carnivores. Oikos, 101: 591-601.

\rf Ingraham, M.W. and S.G. Foster, S.G. (2008). The value of ecosystem services provided by the US National Wildlife Refuge System in the contiguous US. Ecological Economics, 67: 608-618.

\rf Jonsen, I.D., J.M. Flemming, and R.A. Myers. (2005). Robust state–space modeling of animal movement data. Ecology, 86: 2874-2880.

\rf Karasov, W.H. (1992). Daily energy expenditure and the cost of activity in mammals. American Zoology, 32: 238–248.

\rf Kays, R., M. Crofoot, W. Jetz, and M. Wikelski.  (2015).  Terrestrial animal tracking as an eye on life and planet.  Science, 384(6240): aaa2478. 

\rf Kloeden, P.E. and E. Platen. (1992). Numerical Solution of Stochastic Differential Equations. Springer, Berlin.

\rf Knopff, K. H., A. A. Knopff, M. B. Warren, and M. S. Boyce. (2009). Evaluating global positioning system telemetry techniques for estimating cougar predation parameters. Journal of Wildlife Management, 73: 586-597.

\rf Langrock, R., R. King, J. Matthiopoulos, L. Thomas, D. Fortin, and J. Morales.  (2012).  Flexible and practical modeling of animal telemetry data: Hidden Markov models and extensions.  Ecology, 93: 2336-2342.

\rf Laundr{\'e}, J.W., L. Hern{\'a}ndez, and K.B. Altendorf.  (2001).  Wolves, elk, and bison:  Reestablishing the ``landscape of fear'' in Yellowstone National Park.  Canadian Journal of Zoology, 79: 1401-1409.

\rf Lebreton, J.D., K.P. Burnham, J. Clobert, and D.R. Anderson. (1992). Modeling survival and testing biological hypotheses using marked animals: a unified approach with case studies. Ecological Monographs, 62: 67-118.

\rf Leos-Barajas, V., T. Photopoulou, R. Langrock, T.A. Patterson, Y.Y. Watanabe, M. Murgatroyd, and Y.P. Papastamatiou. (2017). Analysis of animal accelerometer data using hidden {M}arkov models.  Methods in Ecology and Evolution, 8: 161-173.

\rf Matthiopoulos, J., J. Fieberg, G. Aarts, H.L. Beyer, J.M. Morales, and D.T. Haydon. (2015). Establishing the link between habitat selection and animal population dynamics. Ecological Monographs, 85: 413-436.

\rf McClintock, B.T., R. King, L. Thomas, J. Matthiopoulos, B.J. McConnell, and J.M. Morales. (2012). A general discrete-time modeling framework for animal movement using multistate random walks. Ecological Monographs, 82: 335-349.

\rf MacFadyen, S., N. Zambatis, A.J. Van Teeffelen, and C. Hui. (2018). Long-term rainfall regression surfaces for the Kruger National Park, South Africa:  A spatio‐temporal review of patterns from 1981 to 2015.  International Journal of Climatology, 38: 2506-2519.

\rf McKellar, A.E., R. Langrock, J.R. Walters, and D.C. Kesler. (2015). Using mixed hidden Markov models to examine behavioral states in a cooperatively breeding bird. Behavioral Ecology, 26: 148-157.

\rf Merkle, J.A., J.R. Potts, and D. Fortin. (2017).  Energy benefits and emergent space use patterns of an empirically parameterized model of memory-based patch selection.  Oikos, 126: 185-195.  

\rf Morales, J.M., D.T. Haydon, J. Frair, K.E. Holsinger, and J.M. Fryxell.  (2004). Extracting more out of relocation data:  Building movement models as mixtures of random walks. Ecology, 85: 2436-2445.

\rf Morales J.M., D. Fortin, J. Frair and E. Merrill.  (2005).  Adaptive models for large herbivore movements in heterogeneous landscapes. Landscape Ecology, 20: 301-316.

\rf Morales J.M., P.R. Moorcroft, J. Matthiopoulos, J.L. Frair, J.K. Kie, R.A. Powell, E.H. Merrill, and D.T. Haydon.  (2010).  Building the bridge between animal movements and population dynamics. Philosophical Transactions of the Royal Society, Series B, 365: 2289-2301.

\rf Naidoo, R., P. Du Preez, G. Stuart-Hill, M. Jago, and M. Wegmann. (2012). Home on the range: factors explaining partial migration of African buffalo in a tropical environment. PLoS one, 7(5), e36527.

\rf Nathan, R., W.M. Getz, E. Revilla, M. Holyoak, R. Kadmon, D. Saltz, and P.E. Smouse. (2008). A movement ecology paradigm for unifying organismal movement research.  Proceedings of the National Academy of Sciences, 105: 19052-19059.

\rf Nussbaum, M. (1978).  Aristotle's De Motu Animalium: Text with Translation, Commentary, and Interpretive Essays.  Princeton University Press, Princeton, New Jersey, USA. 

\rf Oliveira-Santos, L.G.R., J.D. Forester, U. Piovezan, W.M. Tomas, and F.A. Fernandez. (2016). Incorporating animal spatial memory in step selection functions. Journal of Animal Ecology, 85: 516-524.

\rf Pollock, K.H. (1991).  Modeling capture, recapture, and removal statistics for estimation of demographic parameters for fish and wildlife populations:  Past, present, and future.  Journal of the American Statistical Association, 86: 225-238.

\rf Preisler, H.K., A.A. Ager, and M.J. Wisdom. (2013). Analyzing animal movement patterns using potential functions. Ecosphere, 4: 1-13.

\rf Prins, H.H.T. (1996). Ecology and Behaviour of the African Buffalo: Social Inequality and Decision Making. London: Chapman \& Hall.

\rf Prins, H.H.T. and A.R.E. Sinclair. (2013).  Syncerus caffer. In:  The Mammals of Africa. Kingdon, J.S. and M. Hoffmann (eds.). Amsterdam: Academic Press.

\rf Proaktor, G., T. Coulson, and E.J. Milner-Gulland. (2008). The demographic consequences of the cost of reproduction in ungulates. Ecology, 89: 2604-2611.

\rf Radloff, F.G. and J.T. Du Toit. (2004). Large predators and their prey in a southern African savanna: a predator's size determines its prey size range. Journal of Animal Ecology 73: 410-423.

\rf Redfern, J.V., R. Grant, H. Biggs, and W.M. Getz. (2003). Surface‐water constraints on herbivore foraging in the Kruger National Park, South Africa. Ecology, 84: 2092-2107. 

\rf Ryan, S.J., C.U. Knechtel, and W.M. Getz. (2006). Range and habitat selection of African buffalo in South Africa. The Journal of Wildlife Management, 70: 764-776.

\rf Savage, V.M. and G.B. West. (2007). A quantitative, theoretical framework for understanding mammalian sleep.  Proceedings of the National Academy of Sciences, 104: 1051-1056.

\rf Scharf, H.R., M.B. Hooten, B.K. Fosdick, D.S. Johnson, J.M. London, and J.W. Durban. (2016). Dynamic social networks based on movement. Annals of Applied Statistics, 10: 2182-2202.

\rf Scharf, H.R., M.B. Hooten, D.S. Johnson, and J.W. Durban. (2018).  Process convolution approaches for modeling interacting trajectories.  Environmetrics, e2487.

\rf Schaub, M. and F. Abadi. (2011). Integrated population models: A novel analysis framework for deeper insights into population dynamics. Journal of Ornithology, 152: 227-237.

\rf Schick, R.S., S.R. Loarie, F. Colchero, B.D. Best, A. Boustany, D.A. Conde, P.N. Halpin, L.N. Joppa, C.M. McClellan, and J.S. Clark. (2008). Understanding movement data and movement processes:  Current and emerging directions.  Ecology Letters, 11: 1338-1350.

\rf Schick, R.S., S.D. Kraus, R.M. Rolland, A.R. Knowlton, P.K. Hamilton, H.M. Pettis, R.D. Kenney, and J.S. Clark. (2013). Using hierarchical Bayes to understand movement, health, and survival in the endangered North Atlantic right whale. PloS one, 8: e64166.

\rf Shepard, E.L.C., R.P. Wilson, W.G. Rees, E. Grundy, S.A. Lambertucci, and S.B. Vosper. (2013). Energy landscapes shape animal movement ecology. American Naturalist, 182: 298-312.

\rf Sinclair, A.R.E. (1977).  The African Buffalo: A Study of Resource Limitation of Populations.  University of Chicago Press.  Chicago, IL, USA.

\rf Spalinger, D.E. and N.T. Hobbs. (1992).  Mechanisms of foraging in mammalian herbivores:  New models of functional response. The American Naturalist, 140: 325-348.

\rf Stephens, D.W., J.S. Brown, and R. Ydenberg.  (2007).  Foraging Behavior and Ecology.  University of Chicago Press.  Chicago, IL, USA.

\rf Taylor, C.R., N.C. Heglund, and G.M. Maloiy. (1982). Energetics and mechanics of terrestrial locomotion. I. Metabolic energy consumption as a function of speed and body size in birds and mammals. Journal of Experimental Biology, 97: 1-21.

\rf Tshipa, A., H. Valls-Fox, H. Fritz, K. Collins, L. Sebele, P. Mundy, and S. Chamaill\'e-Jammes. (2017). Partial migration links local surface-water management to large-scale elephant conservation in the world's largest transfrontier conservation area. Biological Conservation, 215: 46-50.

\rf Turchin, P. (1998).  Quantitative Analysis of Movement: Measuring and Modeling Population Redistribution in Animals and Plants. Sinauer. Sunderland, Massachusetts, USA.  

\rf Williams, T. M., L. Wolfe, T. Davis, T. Kendall, B. Richter, Y. Wang, C. Bryce, G.H. Elkaim, and C.C. Wilmers. (2014). Instantaneous energetics of puma kills reveal advantage of felid sneak attacks. Science, 346: 81-85.

\rf Wilmers, C.C., B. Nickel, C.M. Bryce, J.A. Smith, R.E. Wheat, and V. Yovovich.  (2015). The golden age of bio‐logging: How animal‐borne sensors are advancing the frontiers of ecology. Ecology, 96: 1741-1753.

\rf Wilmers, C.C., L.A. Isbell, J.P. Suraci, and T.M. Williams. (2017). Energetics-informed behavioral states reveal the drive to kill in {A}frican leopards. Ecosphere, 8: e01850.

\rf Wilson, R.P., M.P.T. Wilson, R. Link, H. Mempel, and N.J. Adams. (1991). Determination of movements of {A}frican penguins {S}pheniscus demersus using a compass system: Dead reckoning may be an alternative to telemetry. Journal of Experimental Biology, 157: 557-564.

\rf Wilson, R., F. Quintana, and V. Hobson. (2012). Construction of energy landscapes can clarify the movement and distribution of foraging animals. Proceedings of the Royal Society, Series B, 279: 975-980.

\rf Winnie, J.A., P. Cross, and W. Getz. (2008). Habitat quality and heterogeneity influence distribution and behavior in African buffalo (\emph{Syncerus caffer}). Ecology, 89: 1457-1468.

\rf Whoriskey, K., M. Auger-M\'eth\'e, C.M. Albertsen, F.G. Whoriskey, T.R. Binder, C.C. Krueger, and J. Mills Flemming. (2017). A hidden Markov movement model for rapidly identifying behavioral states from animal tracks. Ecology and Evolution, 7: 2112-2121.

\rf Zuntz, N. (1897).  Uber den Stoffverbrauch des Hundes bei Muskelarbeit.  Pflugers Archiv, 68: 191-211.

\pagebreak

\pagebreak

\pagebreak

\pagebreak

\pagebreak

\pagebreak

\pagebreak

\clearpage
\section*{Appendix A: Derivation of Aggregated Recharge Process}
In what follows, we show how to obtain the aggregated recharge process $g(t)$ as a function of the physiological recharge functions $g(v,t)$.  Consider the integral of $g(v,t)$ over ${\cal V}$

\begin{align}
  g(t) &= \int_{{\cal V}} g(v,t) dv \,, \\ 
  &= \int_{{\cal V}} g_0(v) + \int_0^t \mathbf{w}'(\boldsymbol\mu(\tau))\boldsymbol\theta(v) d\tau dv \,,  \\ 
  &= g_0 + \int_{{\cal V}}\int_0^t \mathbf{w}'(\boldsymbol\mu(\tau))\boldsymbol\theta(v) d\tau dv \,, \\ 
  &= g_0 + \int_0^t\int_{{\cal V}} \mathbf{w}'(\boldsymbol\mu(\tau))\boldsymbol\theta(v) dv d\tau \,, \label{eq:gt_key} \\ 
  &= g_0 + \int_0^t \mathbf{w}'(\boldsymbol\mu(\tau))\boldsymbol\theta d\tau \,, 
\end{align}
The key aspect of the derivation above in involves the use of Fubini's theorem to change the order of integration in (\ref{eq:gt_key}) which yields the following results  

\begin{align} 
  g_0&=\int_{{\cal V}} g_0(v) dv \,,  \\
  \boldsymbol\theta&=\int_{{\cal V}} \boldsymbol\theta(v) dv \,. 
\end{align} 

\section*{Appendix B:  Analysis of Simulated Data}
\begin{figure}[htp]
\centering
\includegraphics[width=4in]{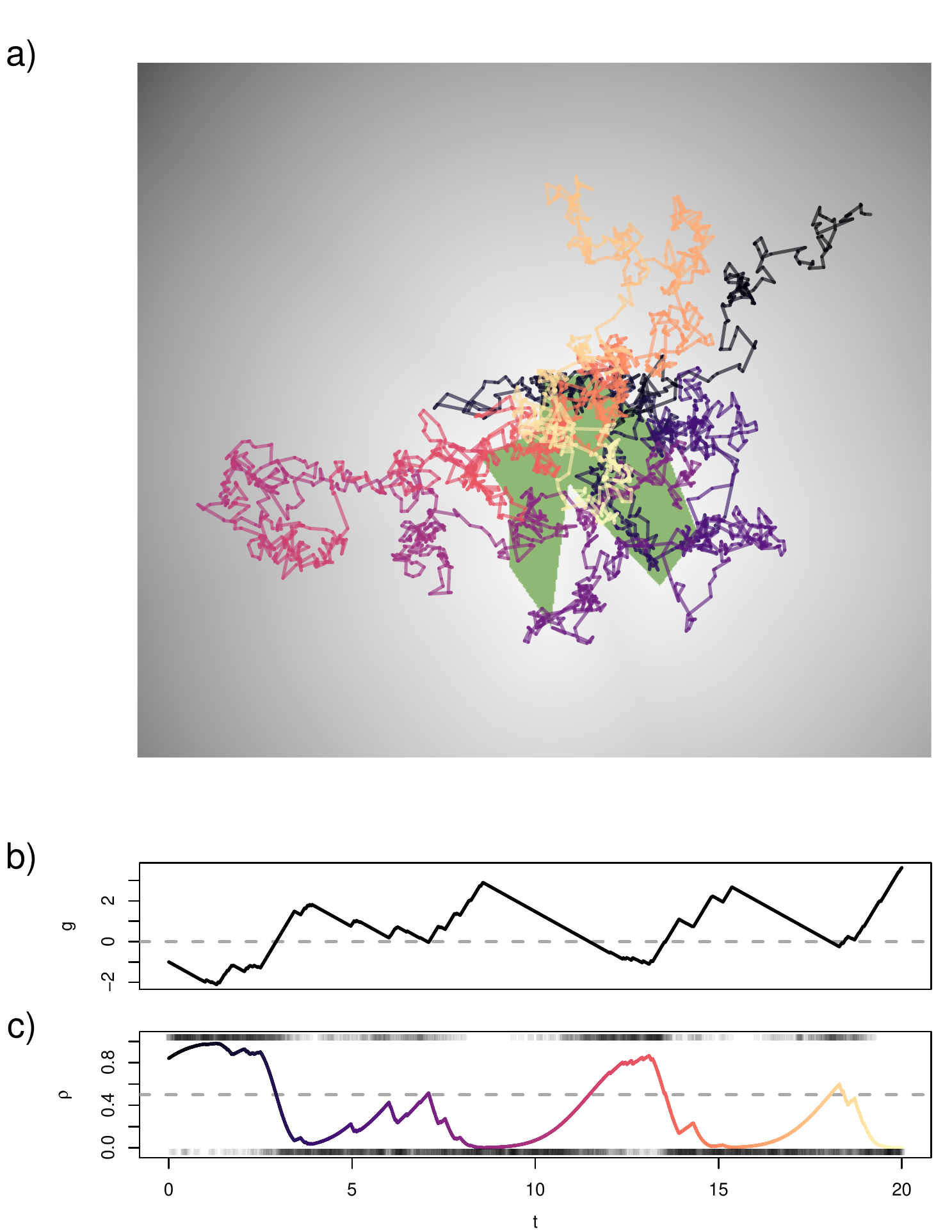}
\caption{a) Simulated trajectory $\boldsymbol\mu(t)$ in two dimensions and b) associated recharge ($g(t)$) and c) decision probability ($\rho(t)$).  The green polygon shown in the middle of (a) represents a patch that corresponds to recharge and the single covariate based on distance to patch is shown in (a) as shading from light (near patch) to dark (far from patch).  The trajectory color corresponds to the time elapsed since the initial point of the study period (increasing from dark blue to yellow).  The simulated decision process $z(t)$ is shown as semitransparent points at the top ($z(t)=1$) and bottom ($z(t)=0$) of the plot for $\rho(t)$ in (c).}
\label{fig:sim_data}
\end{figure}

To demonstrate the hierarchical movement model with recharge dynamics, we simulated a set of telemetry data with $n=500$ observations based on the following specification for the movement and recharge components of the model.  The first model component ${\cal M}_0$, implies our simulated trajectory will arise as a random walk resembling nomadic behavior when the individual is in a charged state.  For the potential function associated with the decision to recharge, we specified $p(\boldsymbol\mu(t),\boldsymbol\beta)=\mathbf{x}'(\boldsymbol\mu(t))\boldsymbol\beta$, implying that the movement of the simulated individual may be influenced by the gradient of a linear combination of our covariates $\mathbf{x}(\boldsymbol\mu)$.  In our simulation, we considered a single spatial covariate $x(\boldsymbol\mu(t))$ defined as the Euclidean distance to a patch polygon in the study area (Figure~\ref{fig:sim_data}a; where the patch itself may be desirable to the individual because it represents a central place, energy resource, etc).  Thus, the potential function for our recharge movement process simplifies to $p(\boldsymbol\mu(t),\beta)=\beta x(\boldsymbol\mu(t))$ (with $\beta=-55658.2$ simulating a trajectory with drift toward the patch when the individual is discharged).  Note that there is no intercept in this potential function because it does not affect the gradient in the movement model. 

We specified the recharge process based on the expression $g(t)$ and using $g_0=-1$ and $\mathbf{w}(\boldsymbol\mu(t))=(1,w(\boldsymbol\mu(t)))'$ where $w(\boldsymbol\mu(t))=1$ if the individual is in the recharge patch (green polygon in Figure~\ref{fig:sim_data}a) at time $t$, and zero otherwise.  For recharge coefficients, we used $\boldsymbol\theta=(\theta_0,\theta_1)'$ with $\theta_0=-1$ and $\theta_1=4$.  This formulation for the recharge function $g(t)$ implies a recharge rate that is three times as fast as the discharge rate, which may be more realistic for certain types of recharge processes (e.g., energetics; Figure~\ref{fig:sim_data}b).  

The remaining parameters used to simulate data were specified as: $\sigma^2_s=10^{-5}$, $\sigma^2_0=0.02$, and $\sigma^2_1=0.03$.  For the hierarchical movement model based on recharge dynamics fit to simulated data, we specified priors for each of the model parameters as: 

\begin{align}
\lb \sigma_s^2 \rb &= \text{IG}(q_s = 2.000122, r_s = 3.000122 \times 10^{-5})
\intertext{Hyperparameters chosen so that mode of prior is $10^{-5}$ and variance is $10^{-2}$.}
\lb \sigma_0^2 \rb &= \text{IG}(q_0 = 2.003556, r_0 = 0.06007113)
\intertext{Hyperparameters chosen so that mode of prior is $0.02$ and variance is $1$.}
\lb \sigma_1^2 \rb &= \text{IG}(q_1 = 2.008019, r_1 = 0.09024058)
\intertext{Hyperparameters chosen so that mode of prior is $0.03$ and variance is $1$.}
\lb \boldsymbol\beta \rb &=
  \text{N}(\boldsymbol\mu_\beta, \boldsymbol\Sigma_\beta), \quad \boldsymbol\mu_\beta = 0, \; \; \boldsymbol\Sigma_\beta = 2.5^{13} \\
\lb g_0 \rb &= \text{N}(\mu_{g_0}, \sigma_{g_0}^2), \quad \mu_{g_0} = 0, \; \; \sigma_{g_0}^2 = 1 \\
\lb \boldsymbol\theta \rb &=
  \text{N}(\boldsymbol\mu_\theta, \boldsymbol\Sigma_\theta), \quad \boldsymbol\mu_\theta = \mathbf{0}, \; \; \boldsymbol\Sigma_\theta = 1000\mathbf{I}_2
\end{align}
\noindent The temporal grid contained $m = 1500$ time points. We standardized all covariates so that the mean slope along the trajectory was approximately 1. We then specified weakly informative zero-mean normal priors for $\boldsymbol\beta$ with standard deviation 50, corresponding to a belief that the contribution made by the covariate to the total displacement of an individual should be less than about 100 spatial units per full temporal unit. On the scale of the untransformed covariate, this corresponds to a standard deviation of approximately $5^6$. For $\boldsymbol\theta$, the prior standard deviation of 31.62 corresponds to a belief that the individual will change from fully charged to fully depleted no more than about 50 times per whole time unit.  

\begin{figure}[htp]
\centering
\includegraphics[width=4in]{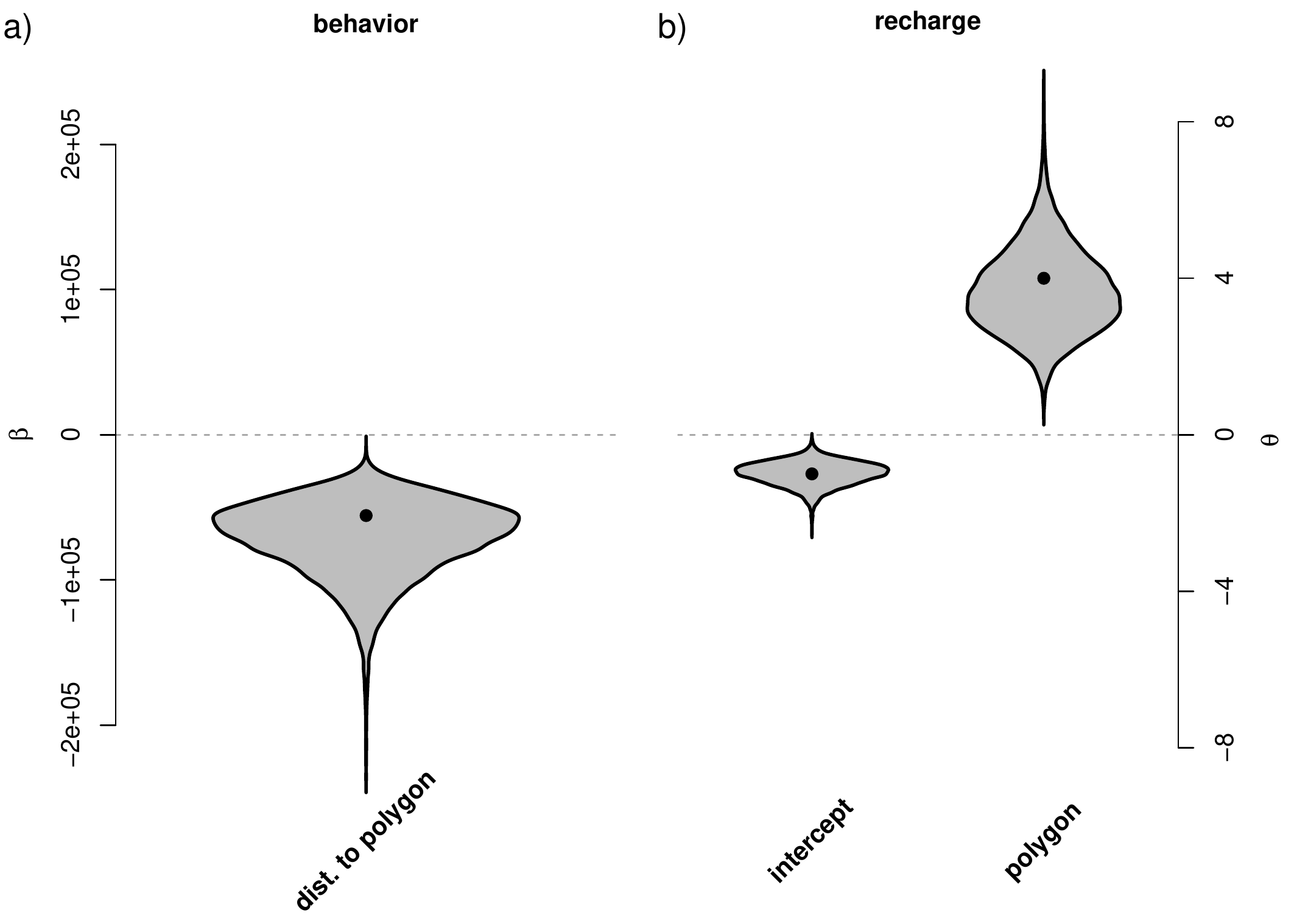}
\caption{Marginal posterior violin plots for a) $\beta_1$ (left), and b) $\theta_0$ (center) and $\theta_1$ (right).  True values used to simulate data shown as black points.}
\label{fig:sim_post_params}
\end{figure}

We fit the recharge-based movement model to the simulated data shown in Figure~\ref{fig:sim_data}.  We also fit the simpler hierarchical models including only ${\cal M}_1$ ($z(t)=1$ for all $t$) and ${\cal M}_0$ ($z(t)=0$ for all $t$) and scored them to assess predictive ability using the approach described in Appendix C.  Cross-validation indicated that we were able to correctly identify the recharge-based movement model as the data generating model when compared to the simpler alternatives based only on ${\cal M}_1$ and ${\cal M}_0$.  The marginal posterior distributions for $\beta$ and $\boldsymbol\theta$ are shown in Figure~\ref{fig:sim_post_params}.  

\begin{figure}[htp]
\centering
\includegraphics[width=6in]{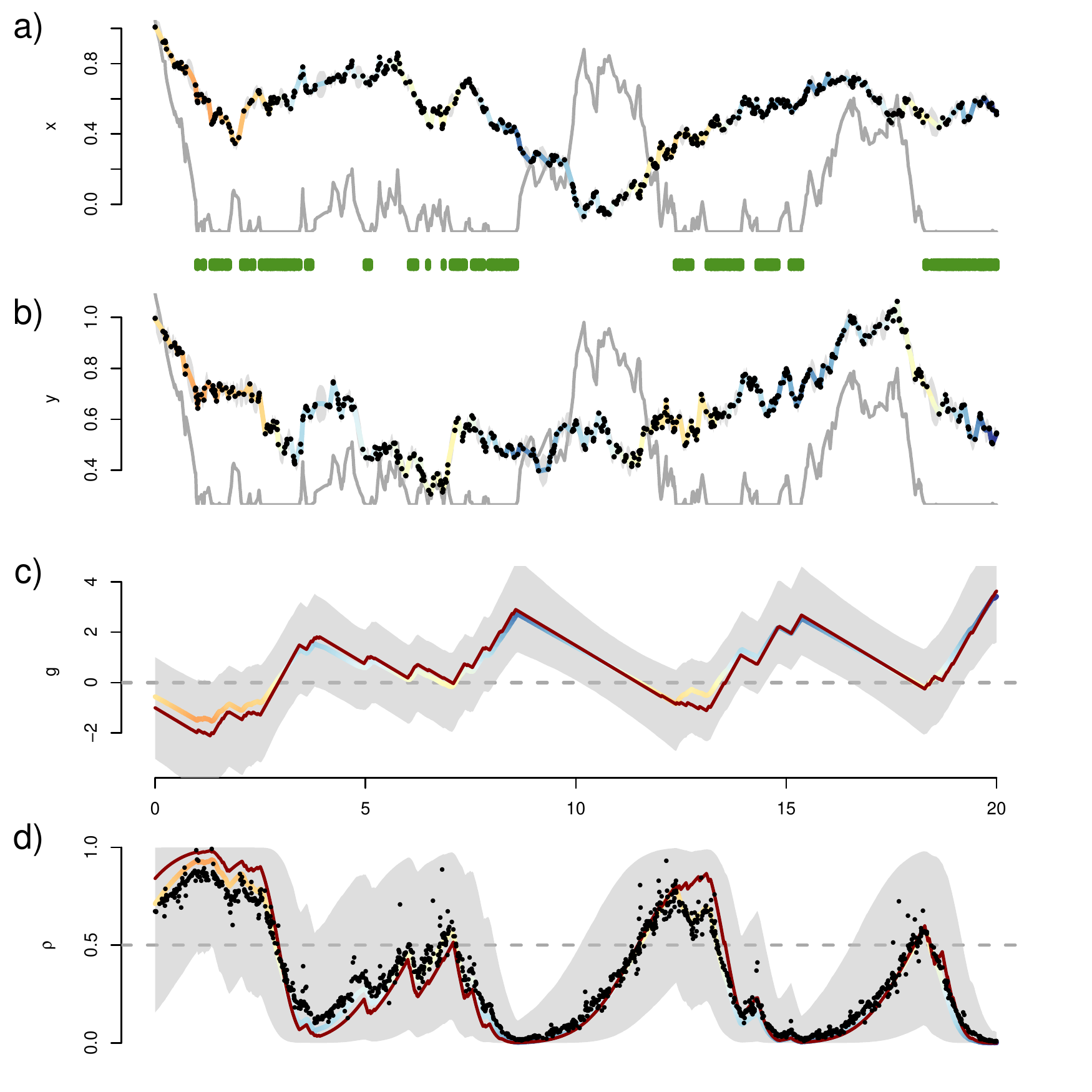}
\caption{Telemetry data (points) and posterior mean trajectory (multicolored line) for the a) easting and b) northing directions.  Gray lines in (a) and (b) represent the value of $x(\boldsymbol\mu(t))$ and green rug plot below (a) represents positions where the recharge function is increasing. The posterior median c) recharge function $g(t)$ and d) associated probability $\rho(t)$ of decision to recharge are shown as multicolored lines with 95\% credible intervals shown in gray ($z(t)$ shown as black points).  True values of $g(t)$ and $\rho(t)$ are shown as red lines.  The color associated with the estimates indicates discharged when orange and charged when blue.}
\label{fig:sim_post_grho}
\end{figure}

The model properly recovers the parameters, indicating that the simulated individual moves toward the patch when needing to recharge and then recharging inside the patch at a faster rate than the discharge outside the patch.  The associated posterior inference for the recharge function $g(t)$ and probability of decision to recharge $\rho(t)$ are shown in Figure~\ref{fig:sim_post_grho}.

The results shown in Figure~\ref{fig:sim_post_grho} indicate that we are able to use the recharge-based movement model to learn about an underlying physiological recharge function using telemetry data alone.  In fact, the posterior median recovers the pattern in the simulated functions (Figure~\ref{fig:sim_post_grho}c--d) quite well, where the uncertainty increases appropriately when the decision probability $\rho(t)$ approaches one half.   

\section*{Appendix C:  Prior Specifications, Scoring, and Computing Details}
For the hierarchical movement model based on recharge dynamics fit to the mountain lion telemetry data, we specified priors for each of the model parameters as: 

\begin{align}
\lb \sigma_s^2 \rb &= \text{IG}(q_s = 4.479787, r_s = 54.79787)
\intertext{Hyperparameters chosen so that mode of prior is $10$ and variance is $100$.}
\lb \sigma_0^2 \rb &= \text{IG}(q_0 = 4.479815, r_0 = 21919260)
\intertext{Hyperparameters chosen so that mode of prior is $4\times 10^6$ and variance is $4\times 10^{12}$.}
\lb \sigma_1^2 \rb &= \text{IG}(q_1 = 4.479815, r_1 =  21919260)
\intertext{Hyperparameters chosen so that mode of prior is $4\times 10^6$ and variance is $4\times 10^{12}$.}
\lb \boldsymbol\beta \rb &=
  \text{N}(\boldsymbol\mu_\beta, \boldsymbol\Sigma_\beta), \quad \boldsymbol\mu_\beta = 0, \; \; \boldsymbol\Sigma_\beta = 2.5\times 10^{7}\mathbf{I}_5 \\
\lb g_0 \rb &= \text{N}(\mu_{g_0}, \sigma_{g_0}^2), \quad \mu_{g_0} = 0, \; \; \sigma_{g_0}^2 =  \\
\lb \boldsymbol\theta \rb &=
  \text{N}(\boldsymbol\mu_\theta, \boldsymbol\Sigma_\theta), \quad \boldsymbol\mu_\theta = \mathbf{0}, \; \; \boldsymbol\Sigma_\theta = 532 \mathbf{I}_5
\end{align}
\noindent The temporal grid contained $m = 600$ time points. We standardized all $\mathbf{X}$ covariates so that the mean slope along the trajectory was approximately 1. We then specified weakly informative zero-mean normal priors for $\boldsymbol\beta$ with standard deviation 5000, corresponding to a belief that the contribution made by any given covariate to the total displacement of an individual should be less than about 10 km per day.  For $\boldsymbol\theta$, a standard deviation of 23.07 corresponds to a belief that the individual will change from fully charged to fully depleted no more than about once per hour.

For the hierarchical movement model based on recharge dynamics fit to the African buffalo telemetry data, we specified priors for each of the model parameters as: 

\begin{align}
\lb \sigma_s^2 \rb &= \text{IG}(q_s = 2.266181, r_s = 3.266181)
\intertext{Hyperparameters chosen so that mode of prior is $1$ and variance is $25$.}
\lb \sigma_0^2 \rb &= \text{IG}(q_0 = 4.479815, r_0 = 493183.4)
\intertext{Hyperparameters chosen so that mode of prior is $9\times 10^4$ and variance is $9\times 10^{8}$.}
\lb \sigma_1^2 \rb &= \text{IG}(q_1 = 4.479815, r_1 = 493183.4)
\intertext{Hyperparameters chosen so that mode of prior is $9\times 10^4$ and variance is $9\times 10^{8}$.}
\lb \boldsymbol\beta \rb &=
  \text{N}(\boldsymbol\mu_\beta, \boldsymbol\Sigma_\beta), \quad \boldsymbol\mu_\beta = 0, \; \; \boldsymbol\Sigma_\beta = 10^4\mathbf{I}_5 \\
\lb g_0 \rb &= \text{N}(\mu_{g_0}, \sigma_{g_0}^2), \quad \mu_{g_0} = 0, \; \; \sigma_{g_0}^2 = 1 \\
\lb \boldsymbol\theta \rb &=
  \text{N}(\boldsymbol\mu_\theta, \boldsymbol\Sigma_\theta), \quad \boldsymbol\mu_\theta = \mathbf{0}, \; \; \boldsymbol\Sigma_\theta = 359\mathbf{I}_5
\end{align}
\noindent The temporal grid contained $m = 1441$ time points.  We standardized all $\mathbf{X}$ covariates so that the mean slope along the trajectory was approximately 1. We then specified weakly informative zero-mean normal priors for $\boldsymbol\beta$ with standard deviation 100, corresponding to a belief that the contribution made by any given covariate to the total displacement of an individual should be less than about 200m per hour.  For $\boldsymbol\theta$, a standard deviation of 18.95, which corresponds to a belief that the individual will change from fully charged to fully depleted no more than about once per hour.

We fit the mountain lion and African buffalo models using 100000 MCMC iterations (discarding the first 50000 as burn-in and thinning the remainder at an interval of 10 iterations) which required approximately 20 hours for the most complex model on a workstation with 3 Ghz 8-core Intel Xeon processor and 64 GB of RAM.   

To score the models, we used the negative log posterior predictive density as a proper predictive score computed using 8-fold cross-validation (Hooten and Hobbs, 2015).  The posterior predictive distribution is analytically intractable (along with the rest of the posterior quantities of interest, as is typical in hierarchical Bayesian models), thus, we estimated it using Monte Carlo integration based on a kernel density estimate of the point-wise posterior predictive density evaluated at the hold out data.  We averaged the resulting quantities across equal-sized folds to arrive at the final score for each model.  Smaller scores indicate better predictive performance.  

The predictive scores for: 1) the mountain lion models are shown in Table~\ref{tab:ml_scores}, for the African buffalo models are shown in Table~\ref{tab:ab_scores}, and 3) for the simulation models are shown in Table~\ref{tab:sim_scores}.
\begin{table}
\centering
\begin{tabular}{lc}
Model & Score \\
\hline
Recharge full & 15.45199 \\
Recharge reduced & 15.38224 \\
${\cal M}_{1,\text{full}}$ & 15.47527 \\
${\cal M}_{1,\text{reduced}}$ & 15.44072 \\
${\cal M}_0$ & 15.42810 \\
\end{tabular}
\caption{Predictive scores for mountain lion models (smaller scores indicate better predictive models). Recharge full: recharge model including all covariates.  Recharge reduced: recharge model including only prey kill area covariates.  ${\cal M}_{1,\text{full}}$: Hierarchical model including all movement covariates assuming that all $z(t)=1$.  ${\cal M}_{1,\text{reduced}}$: Hierarchical model including only prey kill area covariate, assuming that all $z(t)=1$. ${\cal M}_0$:  Hierarchical model with no covariates, assuming that all $z(t)=0$.}
\label{tab:ml_scores}
\end{table}

\begin{table}
\centering
\begin{tabular}{lc}
Model & Score \\
\hline
Recharge full & 13.34667 \\
Recharge reduced & 13.29791 \\
${\cal M}_{1,\text{full}}$ & 13.61633 \\
${\cal M}_{1,\text{reduced}}$ & 13.59157 \\
${\cal M}_0$ & 13.57993 \\
\end{tabular}
\caption{Predictive scores for African buffalo models (smaller scores indicate better predictive models). Recharge full: Recharge model including all covariates. Recharge reduced: recharge model including only surface water covariates.  ${\cal M}_{1,\text{full}}$: Hierarchical model including all movement covariates assuming that all $z(t)=1$. ${\cal M}_{1,\text{reduced}}$: Hierarchical model including only distance to surface water covariate, assuming that all $z(t)=1$.  ${\cal M}_0$:  Hierarchical model with no covariates, assuming that all $z(t)=0$.}
\label{tab:ab_scores}
\end{table}

\begin{table}
\centering
\begin{tabular}{lc}
Model & Score \\
\hline
Recharge & -5.458882 \\
${\cal M}_1$ & -5.447527 \\
${\cal M}_0$ & -5.443709 \\
\end{tabular}
\caption{Predictive scores for simulated data models (smaller scores indicate better predictive models). Recharge: recharge model including recharge region covariates.  ${\cal M}_1$: Hierarchical model including recharge region covariates assuming that all $z(t)=1$.  ${\cal M}_0$:  Hierarchical model with no covariates assuming that all $z(t)=0$.}
\label{tab:sim_scores}
\end{table}

\section*{Appendix D: Pseudo-code for obtaining draws from posterior distribution}

To fit the model we used an MCMC algorithm with the following updates for model parameters:
\begin{enumerate}
\item Update $\sigma_s^2 \sim \lb \sigma_s^2 | \cdot \rb =
  \text{IG}(q_s + n, r_s + \frac{1}{2} \sum_{i = 1}^n \sum_{d=1}^2 \lp s(t_i, d) - \mu(t_i, d)\rp^2 )$ where $d$ indexes the two spatial dimensions (e.g., longitude and latitude).
\item Define a fine grid of time points $\mathcal{T}$ over the study interval $\lb 0, T \rb$ of size $m$ such that $\mathcal{T} \supset \mathcal{T}_{obs}$, where $\mathcal{T}_{obs}\equiv \{t_1,\ldots,t_i,\ldots,t_n\}$ is the set of observation time points. Let $\Delta t_j = t_j - t_{j-1}$.
  \begin{enumerate}
  \item Update

    \begin{align}
    &\boldsymbol\mu(t_1) \sim \lb \boldsymbol\mu(t_1) | \cdot \rb \\
    &\qquad \propto \lb \boldsymbol\mu(t_2) | \boldsymbol\mu(t_1), \sigma_0^2, \sigma_1^2, \boldsymbol\beta, z(t_2) \rb
      \lb \mathbf{s}(t_1) | \boldsymbol\mu(t_1), \sigma_s^2 \rb^{\bone{t_1 \in \mathcal{T}_{obs}}} \prod_{j=2}^m [z(t_j)|g_0,\boldsymbol\theta]
    \end{align}
    using a Metropolis random walk with an adaptively tuned bivariate Gaussian proposal distribution.  Note that we have suppressed notation for the position process $\boldsymbol\mu(t)$ in the conditional distribution for $z(t_j)$ to simplify the expressions.   
  \item For $j = 2, \dots, m - 1$ update

  \begin{align}
    &\boldsymbol\mu(t_j) \sim \lb \boldsymbol\mu(t_j) | \cdot \rb \propto \\
    &\qquad \lb \boldsymbol\mu(t_{j + 1}) | \boldsymbol\mu(t_j), \sigma_0^2, \sigma_1^2, \boldsymbol\beta, z(t_{j+1}) \rb
    \lb \boldsymbol\mu(t_j) | \boldsymbol\mu(t_{j-1}), \sigma_0^2, \sigma_1^2, \boldsymbol\beta, z(t_j) \rb \\
    &\qquad \times \lb \mathbf{s}(t_j) | \boldsymbol\mu(t_j), \sigma_s^2 \rb^{\bone{t_j \in \mathcal{T}_{obs}}} \prod_{l=j+1}^m [z(t_l)|g_0,\boldsymbol\theta]
  \end{align}
  using a Metropolis random walk with an adaptively tuned bivariate Gaussian proposal distribution.
  \item Update

  \begin{align}
    &\boldsymbol\mu(t_m) \sim \lb \boldsymbol\mu(t_m) | \cdot \rb \\
    &\qquad \propto \lb \boldsymbol\mu(t_m) | \boldsymbol\mu(t_{m-1}), \sigma_0^2, \sigma_1^2, \boldsymbol\beta, z(t_m) \rb
    \lb \mathbf{s}(t_{m}) | \boldsymbol\mu(t_{m}), \sigma_s^2 \rb^{\bone{t_m \in \mathcal{T}_{obs}}}
    \end{align}
 using a Metropolis random walk with an adaptively tuned bivariate Gaussian proposal distribution.
  \end{enumerate}
  The conditional distributions for $\boldsymbol\mu(t_{j+1})$ are given by

  \begin{align}
    & \lb \boldsymbol\mu(t_{j+1}) | \boldsymbol\mu(t_j), \sigma_0^2, \sigma_1^2, \boldsymbol\beta, z(t_{j+1})\rb = \\
    &\qquad \text{N}\!\lp \boldsymbol\mu(t_j) - z(t_{j+1})\nabla \mathbf{x}'\lp\boldsymbol\mu(t_j)\rp\boldsymbol\beta\Delta t_{j+1},
      \lp(1-z(t_{j+1}))\sigma_0^2 + z(t_{j+1})\sigma_1^2\rp\Delta t_{j+1} \mathbf{I}_2\rp.
  \end{align}
  \item For $j = 1, \dots, m$ update

  \begin{align}
  z(t_j) &\sim \lb z(t_j) | \cdot \rb =
  \text{Bern}\lp\frac{\pi^{(1)}_j}{\pi^{(0)}_j + \pi^{(1)}_j}\rp
  \intertext{where}
  \pi_j^{(1)} &= \lp 1 - \Phi(g(t_j)) \rp
    \lb \boldsymbol\mu(t_{j+1}) | \boldsymbol\mu(t_j), \sigma_0^2, \sigma_1^2, \boldsymbol\beta, z(t_j) = 1 \rb \\
  \pi_j^{(0)} &= \Phi(g(t_j))
    \lb \boldsymbol\mu(t_{j+1}) | \boldsymbol\mu(t_j), \sigma_0^2, \sigma_1^2, \boldsymbol\beta, z(t_j) = 0 \rb \;,
  \end{align}
  where the recharge function is approximated as $g(t_j)=g_0+\sum_{l=1}^{j-1} \Delta t_l \mathbf{w}'(\boldsymbol\mu(t_l))\boldsymbol\theta$.  Note that the full-conditional distributions for each $z(t_j)$ are conditionally independent. Therefore, these parameters need not be updated serially.
  \item Update

  \begin{align}
    &\sigma_0^2, \sigma_1^2, \boldsymbol\beta \sim \lb \sigma_0^2, \sigma_1^2, \boldsymbol\beta | \cdot \rb \\
    &\qquad \propto \lp \prod_{j=2}^m \lb \boldsymbol\mu(t_j) | \boldsymbol\mu(t_{j-1}), \sigma_0^2, \sigma_1^2, \boldsymbol\beta, z(t_j) \rb \rp
      \lb \boldsymbol\beta\rb
      \lb \sigma_0^2 \rb \lb \sigma_1^2 \rb
  \end{align}
  as a block using a Metropolis random walk with an adaptively tuned multivariate Gaussian proposal distribution.
  \item Update

  \begin{align}
    &g_0, \boldsymbol\theta \sim \lb g_0, \boldsymbol\theta | \cdot \rb \\
    &\qquad \propto \prod_{j=1}^{m} \lb z(t_j) | g_0, \boldsymbol\theta \rb
      \lb g_0 \rb
      \lb \boldsymbol\theta \rb
  \end{align}
    as a block using a Metropolis random walk with an adaptively tuned multivariate Gaussian proposal distribution.
\end{enumerate}

\section*{Appendix E: Model extension to include auxiliary data}
\begin{figure}[htp]
\centering
\includegraphics[width=1in]{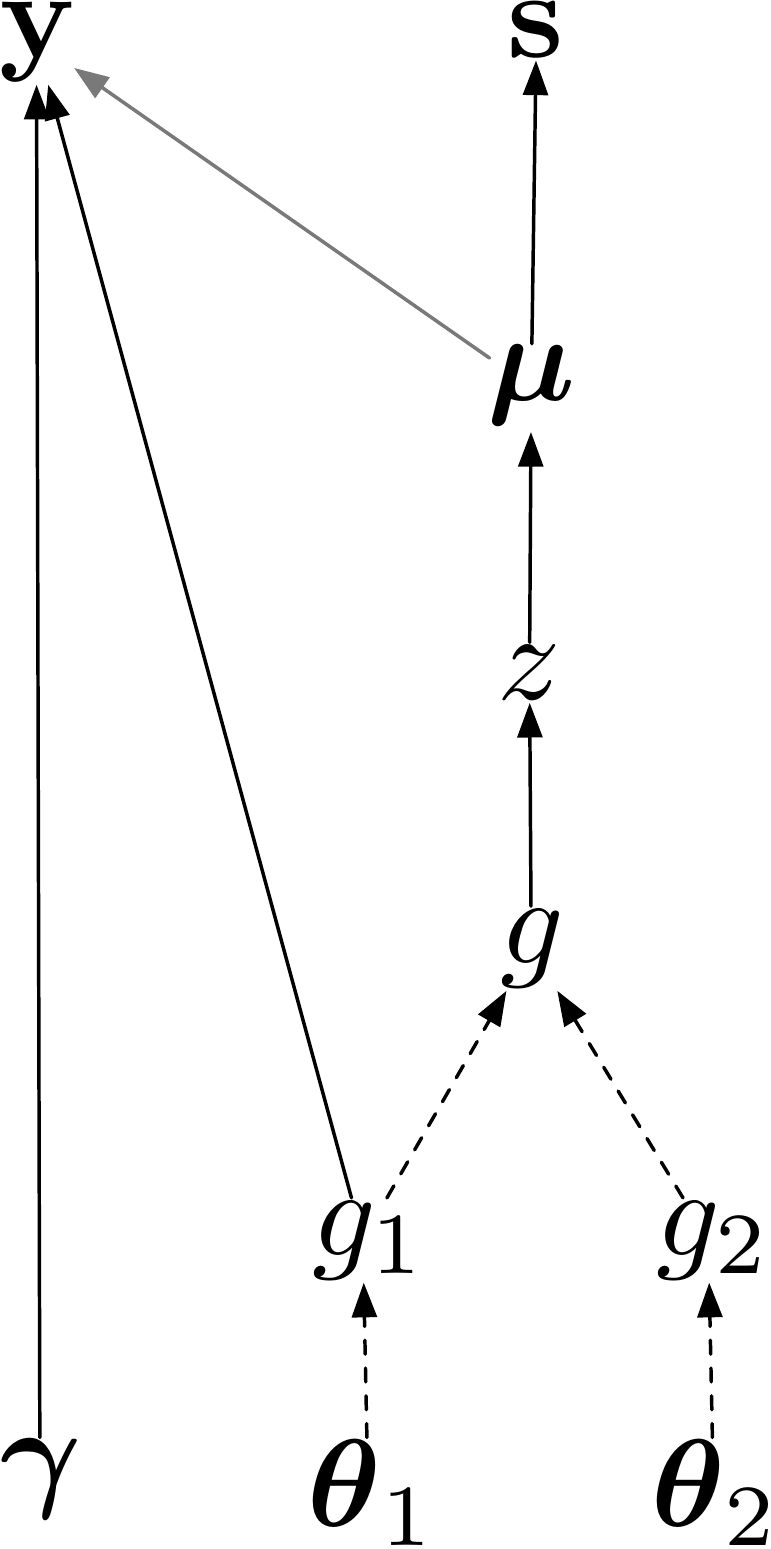}
\caption{Directed acyclic graph (DAG) associated with the recharge-based movement model based on both positional telemetry data ($\mathbf{s}$) and auxiliary data ($\mathbf{y}$).  By convention, solid arrows denote stochastic model dependencies and dashed arrows represent deterministic relationships in the model.  Note that the edge between $\mathbf{y}$ and $\boldsymbol\mu$ is shown in gray because it may or may not exist depending on how the model is specified.}
\label{fig:dag}
\end{figure}

To incorporate auxiliary data in the hierarchical model framework, we retain the model structure for the recharge-based movement model and assume we also have auxiliary data represented by $\mathbf{y}(t_l)$ at time $t_l$ for $l=1,\ldots,L$ observations during the study period.  For example, the $q\times 1$ vectors $\mathbf{y}(t_l)$ could represent a set of $q$ accelerometer measurements recorded at time $t_l$.  Depending on the auxiliary data source and support, we formulate the auxiliary data model generally as 
\begin{equation}      
 \mathbf{y}(t_l) \sim [\mathbf{y}(t_l)|g_1(t_l),\boldsymbol\gamma] \,,  
\end{equation}      
\noindent for $l=1,\ldots,L$ and where $\boldsymbol\gamma$ represents a set of auxiliary data parameters we seek to learn about.  The function $g_1(t_l)$ represents the portion of the aggregated physiological recharge function $g(t_l)$ that relates to the observed auxiliary data $\mathbf{y}(t_l)$.  Thus, the aggregated physiological recharge function $g(t)$ can be partitioned into two components $g(t)=g_1(t)+g_2(t)$, where $g_2(t)$ represents the remainder of physiological processes not measured by $g_1(t)$.  This recharge formulation implies that we have two aggregated physiological recharge functions

\begin{align}   
  g_1(t)&= g_{1,0} + \int_0^t \mathbf{w}_1'(\boldsymbol\mu(\tau))\boldsymbol\theta_1 d\tau \,, \label{eq:g1} \\ 
  g_2(t)&= g_{2,0} + \int_0^t \mathbf{w}_2'(\boldsymbol\mu(\tau))\boldsymbol\theta_2 d\tau \,, \label{eq:g2}  
\end{align}   
\noindent  associated with measured and unmeasured physiological space, respectively.  Each aggregated recharge function is comprised of its own set of parameters we may seek to learn about.  Thus, the full integrated movement model can be visualized by the directed acyclic graph (DAG) shown in Figure~\ref{fig:dag}.  In the DAG, we indicate a possible additional relationship between the auxiliary data ($\mathbf{y}$) and movement process ($\boldsymbol\mu$) using a gray arrow.  

\end{document}